\newcommand\percc{\ifmmode{\rm cm^{-3}}\else{cm$^{-3}$}\fi}
\newcommand\kms{\km\,\pers}
\newcommand\cmsq{\ifmmode{\rm cm^2}\else{cm$^2$}\fi}
\newcommand\coldv{\ifmmode{\cmtwo\,\km^{-1}\,{\rm s}} 
	\else {\cmtwo\,\km$^{-1}\,$s}\fi}
\newcommand\cmtwo{\ifmmode{\rm cm^{-2}}\else{cm$^{-2}$}\fi}
\newcommand\cm{\ifmmode{\rm cm}\else{cm}\fi}
\newcommand\km{\ifmmode{\rm km}\else{km}\fi}
\newcommand\hz{\ifmmode{\rm Hz}\else{Hz}\fi}
\newcommand\mhz{\ifmmode{\rm MHz}\else{MHz}\fi}
\newcommand\ghz{\ifmmode{\rm GHz}\else{GHz}\fi}
\newcommand\pers{\ifmmode{\rm s^{-1}}\else{s$^{-1}$}\fi}
\newcommand\kkms{\ifmmode{\rm K \kms}\else{K \kms}\fi}

\newcommand\pmin{\farcm}

\newcommand\lsim{\la}
\def\ee#1{\ifmmode{\times 10^{#1}}\else{$\times 10^{#1}$}\fi}
\newcommand\jon{\ifmmode{J\!=\!1\!\rightarrow\!0}\else{$J\!=\!1\!\rightarrow\!0$}\fi}
\newcommand\jtw{\ifmmode{J\!=\!2\!\rightarrow\!1}\else{$J\!=\!2\!\rightarrow\!1$}\fi}
\newcommand\jth{\ifmmode{J\!=\!3\!\rightarrow\!2}\else{$J\!=\!3\!\rightarrow\!2$}\fi}
\newcommand\jfo{\ifmmode{J\!=\!4\!\rightarrow\!3}\else{$J\!=\!4\!\rightarrow\!3$}\fi}
\newcommand\jfi{\ifmmode{J\!=\!5\!\rightarrow\!4}\else{$J\!=\!5\!\rightarrow\!4$}\fi}
\newcommand\jsi{\ifmmode{J\!=\!6\!\rightarrow\!5}\else{$J\!=\!6\!\rightarrow\!5$}\fi}
\newcommand\jse{\ifmmode{J\!=\!7\!\rightarrow\!6}\else{$J\!=\!7\!\rightarrow\!6$}\fi}
\newcommand\jei{\ifmmode{J\!=\!8\!\rightarrow\!7}\else{$J\!=\!8\!\rightarrow\!7$}\fi}
\newcommand\maff{\ifmmode{\rm Maffei~2}\else{Maffei~2}\fi}
\newcommand\msun{\ifmmode{M_\odot}\else{$M_\odot$}\fi}
\newcommand\rsun{\ifmmode{R_\odot}\else{$R_\odot$}\fi}
\newcommand\lsun{\ifmmode{L_\odot}\else{$L_\odot$}\fi}
\newcommand\lfir{\ifmmode{L_{FIR}}\else{$L_{FIR}$}\fi}
\newcommand\vlsr{\ifmmode{V_{LSR}}\else{$V_{LSR}$}\fi}
\newcommand\tmb{\ifmmode{T_{mb}}\else{$T_{mb}$}\fi}
\newcommand\tx{\ifmmode{T_{ex}}\else{$T_{ex}$}\fi}
\newcommand\tdv{\ifmmode{\int\nolimits\tmb dv}\else{$\int\nolimits\tmb dv$}\fi}
\newcommand\tas{\ifmmode{T_A^*}\else{$T_A^*$}\fi}
\newcommand\tsys{\ifmmode{T_{sys}}\else{$T_{sys}$}\fi}
\newcommand\nhh{\ifmmode{n_\htwo}\else{$n_\htwo$}\fi}
\newcommand\ncrit{\ifmmode{n_{crit}}\else{$n_{crit}$}\fi}
\newcommand\rat{\ifmmode{\cal R}\else{${\cal R}$}\fi}

\newcommand\gray{\ifmmode{\gamma{\rm -ray}}\else{$\gamma$-ray}\fi}
\newcommand\grays{\ifmmode{\gamma{\rm -rays}}\else{$\gamma$-rays}\fi}

\newcommand\cii{C~{\small II}}

\newcommand\htwo{\ifmmode{{\rm H}_2}\else{H$_2$}\fi}
\newcommand\hcop{\ifmmode{\rm HCO}^+\else{HCO$^+$}\fi}
\newcommand\ihcop{\ifmmode{\rm H^{13}CO^+}\else{H$^{13}$CO$^+$}\fi}
\newcommand\ihcn{\ifmmode{\rm H^{13}CN}\else{H$^{13}$CN}\fi}
\newcommand\ico{\ifmmode{\rm ^{13}CO}\else{$^{13}$CO}\fi}
\newcommand\co{\ifmmode{\rm ^{12}CO}\else{$^{12}$CO}\fi}
\newcommand\ics{\ifmmode{\rm C^{34}S}\else{C$^{34}$S}\fi}

\documentclass[12pt,preprint]{aastex}

\begin{document}

\title{A Mapping Survey of the \ico\ and \co\ Emission in Galaxies}
\shorttitle{Mapping \ico\ and \co\ in Galaxies}

\author{Timothy A. D. Paglione$^{1,2,3}$, W. F. Wall$^4$, Judith
S. Young$^{1,2}$, Mark H. Heyer$^{1,2}$, Michael Richard$^2$, Michael
Goldstein$^2$, Zeke Kaufman$^2$, Julie Nantais$^2$, Gretchen Perry$^2$}
\email{paglione@umich.edu}

\altaffiltext{1}{Five College Radio Astronomy Observatory}
\altaffiltext{2}{Department of Astronomy, Lederle Graduate
Research Center, University of Massachusetts, Amherst, MA 01003}
\altaffiltext{3}{Current address: Department of Astronomy, University
of Michigan, 500 Church St., Ann Arbor, MI 48109-1090}
\altaffiltext{4}{Instituto Nacional de Astrof\'{\i}sica, Optica
y Electr\'onica, Apartado Postal 216 y 51, 72000 Puebla, Pue.,
M\'exico}

\begin{abstract}


We present spectra of the extended \co\ and \ico\ \jon\ emission along
the major axes of 17 nearby galaxies.  Spatial variations in the
ratio of CO and \ico\ integrated intensities, \rat, are found in
nearly every galaxy observed.  There is an overall variation in \rat\
of 20--40\%\ from the inner 2 kpc to the disk.  Roughly a third of the
survey galaxies have such gradients in \rat\ detected above the
2$\sigma$ confidence level.  Though some galaxies show a lower central
value of \rat, on average \rat\ inside 2 kpc is 10--30\%\ higher than
\rat\ outside of 2 kpc.  The average CO/\ico\ intensity ratio within
the central 2 kpc of the survey sources is $11.6 \pm 0.4$ (based on
the noise) $\pm 1.5$ (based on systematic uncertainties estimated from
daily variations in CO and \ico\ intensities).  The 1$\sigma$
dispersion in \rat\ between galactic nuclei of 4.2 is also quite
large.  The average value of \rat\ outside 2 kpc is $9.8 \pm 0.6 \pm
1.2$ with a standard deviation of 4.5.

An increase in the CO/\ico\ intensity ratio from disk to nucleus may
imply that the conversion factor between CO intensity and \htwo\
column density, $X$, is lower in galactic nuclei.  Also variations in
physical conditions, most notably the gas kinetic temperature, affect
both \rat\ and $X$.  Abundance variations probably do not cause the
gradient in \rat, though we do not rule out a decrease in effective
cloud column densities in galactic nuclei possibly caused by
destructive starburst superwinds.  A modest rise in temperature (less
than a factor of 2 or 3) from outside a 2 kpc radius towards the
nucleus can easily account for the observed gradient.  These results
support previous work implying that $X$ is lower in the center of the
Milky Way and probably most galactic nuclei.  Therefore calculating
\htwo\ masses using the standard Galactic X-factor, especially within
the central few kpc of galaxies, overestimates the true mass by
factors of a few.  The standard X-factor still appears to be
appropriate for galactic disks.

\end{abstract}

\keywords{galaxies: ISM, starburst --- ISM: clouds, molecules}

\section{Introduction}

Central to a complete picture of galaxy evolution is understanding
star formation over large scales.  Given that stars form in molecular
gas, the evolution of a galaxy depends on its gas distribution.  The
main constituent of this gas is molecular hydrogen, which, due to its
symmetry, has no dipole transitions between low-lying levels (i.e., no
permitted rotational transitions).  Consequently, the rotational lines
of carbon monoxide and its isotopic variants (e.g., \ico\ and
C$^{18}$O) have been used to study molecular gas in galaxies.  The
millimeter-wave lines of CO are the brightest among non-masing 
molecular transitions.  In particular, the \jon\ line of \co\
(hereafter written simply as CO) has been invaluable for determining
the distribution of molecular gas column densities within the Milky
Way and other galaxies \citep{dame, fcrao.cosurvey}.

These distributions are inferred assuming that the molecular gas column
density, $N_\htwo$, is traced by the velocity-integrated radiation
temperature (Rayleigh-Jeans brightness temperature), $I({\rm CO}\,
\jon)\equiv\int T_R dv$.  Though the CO emission is usually optically
thick and consequently may not reliably trace column densities, the
ratio $N_\htwo/I_{CO}$, or ``X-factor,'' is found to be roughly
constant on the size scales of many parsecs, with a value of $X =
1.6$\ee{20}\,\cmtwo\,(\kkms)$^{-1}$ in the Galactic disk
\citep{dame, hunter}.  The conventional explanation of how
the optically thick CO line consistently measures column density
requires that giant molecular clouds are in self-gravitational
equilibrium and that the mean density and temperature do not
significantly vary from cloud to cloud \citep{vir.xco, sakamoto}.
Despite these results, there is evidence that $X$ is 2--10 times lower
than the standard value in the central few hundred parsecs of the
Galaxy \citep{sod.gc, dahmen}.  It may also increase by 2--4 times the
standard value in the outer Galaxy \citep{grad.xco, sod.outer,
carpenter}.  Nonetheless, a secondary measure of gas column density
is required to gauge any variation of $X$ with galactocentric radius.

There are suggestions in the literature both for and against the
X-factor varying with position in external galaxies, often from
observations of the rotational lines of \ico, a rarer isotopic variant
of the more commonly observed CO.  Since \ico\ is 30--70 times less
abundant than CO \citep{langer.penzias}, its rotational lines are
usually optically thin on parsec scales.  Therefore, assuming similar
molecular gas physical conditions, observations of \ico\ can give
reliable molecular gas column densities, and any variation in the
ratio of CO and \ico\ \jon\ integrated intensities, \rat, can then
test whether $X$ varies systematically within galaxies.
\citet{rickard.blitz} claim that this ratio is up to a factor of 5
higher in the central 1$'$ than in the disks of six galaxies, implying
that $X$ is up to 5 times smaller in galaxy centers.  In contrast, the
CO/\ico\ intensity ratios observed by \citet{young.sanders} and
\citet{sage.isbell} show no significant differences between the
centers of spiral galaxies and their disks.  However, any study of
molecular gas column densities that uses \ico\ lines must address the
uncertainty in assumptions such as the \ico\ abundance, the gas
kinetic temperature, and whether local thermodynamic equilibrium (LTE)
applies.  Also, \rat\ is a sensitive function of optical depth.
Hence, any observed variation, such as found by \citet{rickard.blitz},
could be attributed to variations in the \ico\ abundance or
excitation, rather than a spatial variation in $X$.  Though separating
these variations from other effects requires observations of other
lines (e.g., \ico\ \jth) and isotopic variants (e.g., C$^{18}$O), a
gradient in \rat\ would indicate an important variation in the
physical or chemical properties of the molecular medium of a galaxy.

In this paper, we present the results of a survey of the \jon\
lines of \ico\ and CO along the major axes of 17 galaxies (see
Table~\ref{tab.sources}).  Unlike previous surveys, systematic
uncertainties between intensities at different positions are reduced
by using a receiver array.  The array also permits efficient
fully-sampled mapping of many positions in a galaxy.

\section{Observations and Data Reduction}\label{obs}

The data were obtained between 1998 December and 2000 May at the Five
College Radio Astronomy Observatory (FCRAO) 14~m telescope.  We
observed the CO and \ico\ \jon\ lines (115.27 and 110.20 GHz) with the
MMIC-based, $4\times4$-element array, SEQUOIA \citep{sequoia}.  The
beam FWHM at 115 GHz was 45$''$, and the main beam efficiency was
$\sim 0.5$.  The pixel spacing on the sky was 1\pmin476, or roughly
two full beam widths.  System temperatures were 130--400 K at 110 GHz,
and 200--1200 K at 115 GHz.  We used the facility broad-band
filterbank, comprised of 64 channels of 5 MHz width, which yielded a
velocity resolution of 13 \kms\ at 115 GHz.

The observed galaxies are the brightest and most extended from the
FCRAO Extragalactic CO Survey of \citet{fcrao.cosurvey} that are
primarily within the R.A. range of 9--16 hours
(Table~\ref{tab.sources}).  The array was aligned with the major axis
of each galaxy, and moved from the edge of the CO disk across its full
extent, with measurements taken at half-pixel (roughly half-beam)
spacing.  Thus each map position was observed with all four receivers
of a row.  Fully sampled maps were not made along the minor axes.
Rather, off-axis data were obtained from the rows of the array
parallel to the major axis, offset by $\pm 1\pmin476$ and $\pm
2\pmin952$.

Except for NGC 253, which has a low declination, we observed above an
elevation of 30\degr\ to avoid large gain variations.  We also
observed below an elevation of 75\degr\ to avoid any possible tracking
errors.  The pointing and focus were checked regularly on SiO maser
sources and Jupiter, and calibration was done every several minutes
using the chopper wheel method.  A linear baseline was removed from
all spectra.  Those requiring higher order fits were discarded.
Spectra with r.m.s.\ noise levels significantly different from those
expected given the system temperature were also discarded.  These data
typically resulted from poor calibrations or non-linear baselines
caused by quickly varying conditions.  The filtered data were then
coadded, weighted by the noise ($1/\sigma^2$).  Therefore the largest
calibration offsets, caused by low elevation and poor weather,
contributed less to the overall results.  Relative gain variations
between pixels were well below 10\%, and since each map position was
observed by all four pixels in a row, this uncertainty was further
reduced.

To assess systematic uncertainties, the line integrated intensities
were monitored for day-to-day variations (mostly at the central
position).  The weighted average CO and \ico\ intensities on any given
day varied from the total season's weighted average by 5--25\%\ and
15--50\%, respectively.  Any data with obvious offsets in pointing
(evident as markedly different spectral line shapes), or calibration
(anomalous line and noise temperatures) were discarded.  In sum, the
{\em relative spatial variations} in the CO and \ico\ intensities, and
their ratio, are very well determined for each galaxy.  However, the
systematic uncertainties in \rat, which are important for comparing
the intensity ratios between galaxies, are roughly 15--60\%\
(Table~\ref{tab.rat}).  Most of these systematic variations are likely
due to slight pointing offsets along the minor axis, which are
difficult to detect in low signal-to-noise spectra, and do not
strongly affect the line shapes.  Periodic observations of Jupiter
indicate that the uncertainties due to variations in calibration
contribute no more than 15--20\%\ to the systematic uncertainty in
line intensity.

\section{Results}

Emission from both CO and \ico\ is detected at nearly every mapped
position along the major axes of these galaxies.  Off-axis emission $>
1\pmin5$ away from the major axis, is detected only in the less
inclined galaxies IC 342, M51 and NGC 6946.  Due to the low
signal-to-noise ratio of the \ico\ spectra off-axis, we do not include
those data in the analysis.  Figure~\ref{fig.spectra} shows the
spectra along the major axes of the survey galaxies.  The similarity
of the CO and \ico\ line shapes with position demonstrates the good
relative pointing (better than 5$''$) between the observations.  The
line integrated intensities ($\int\tas dv$) and CO/\ico\ intensity
ratios (\rat) as functions of position along the major axis are shown
in Figure~\ref{fig.profiles}.

The unweighted average of the CO/\ico\ intensity ratio in the beam
centered on the nucleus is $11.5 \pm 0.3 \pm 1.4$, given the
statistical and systematic uncertainties, respectively
(Table~\ref{tab.rat}).  Statistical uncertainties are based on the
r.m.s.\ noise, line width and channel width, and systematic
uncertainties are estimated as described at the end of
\S~\ref{obs}. The standard deviation in \rat\ from one galaxy center
to another is 5.4.  The weighted average central ratios are $9.0 \pm
0.2$ and $7.7\pm 0.6$ based on the statistical and systematic
uncertainties, respectively.  The central ratios for M51 and M82
deviate significantly from these values.  The high ratio of $27\pm 8$
for M82 is consistent with previous studies \citep{stark.m82,
young.scoville, loiseau.m82, kikumoto}.  However, the low ratio of
$5.4\pm 1.0$ for M51 is nearly a factor of 2 below some previous
observations \citep{young.sanders, gb.m51}.  The individual CO and
\ico\ data for M51 were taken over several widely separated days, and
cover a large range in elevation, weather conditions, and times of
day.  Despite the varied conditions, the data sets agree well with one
another (the estimated systematic uncertainties are 13 and 14\%\ for
CO and \ico, respectively, yielding a 19\%\ uncertainty in \rat).  The
relative pointing was consistent, judging by the line shapes and
radial intensity profiles.  Therefore, we do not correct our values
for M51 for any possible calibration offset, though we will draw no
strong conclusions from the low ratio.  We point out that our
observing technique still results in a very low relative uncertainty
in the spatial variation in \rat\ within the disk of M51.

Figure~\ref{fig.profiles} indicates that nearly all the galaxies show
some significant variation in \rat\ with position.  The most prevalent
trend is for \rat\ to drop with galactocentric distance.  The gradient
in \rat\ is more apparent and significant after the line intensities
are binned to make the uncertainty in \rat\ more uniform
(Figure~\ref{fig.rbin}).  Here non-detections have been excluded.  The
calculated uncertainties in these average ratios are multiplied by
$\sqrt{2}$ to account for the interdependence of the Nyquist sampled
data.  This correction is conservative and may overestimate the error
by about 20\%\ for small bin sizes.  Table~\ref{tab.rat} and
Figure~\ref{fig.nucdisk} compare the weighted average values of \rat\
inside and outside of various radii.  Figure~\ref{fig.nucdisk} shows
that for many galaxies, \rat\ drops by as much as a factor of 2
between the central and outer disk.  On average, \rat\ within a radius
of 2 kpc is 40\%\ higher than the disk value.  No correlations are
found between distance and \rat\ nor its gradient
(Figure~\ref{fig.nucdisk}).

Several galaxies show a significant change in the gradient in \rat\
with the size of the central bin (Figure~\ref{fig.drat}).  Note that
typically the change in \rat\ does not depend strongly on the central
bin size, so the adopted distance to a galaxy is not very important.
For those with significant changes, the bin size simply scales with
distance, and intermediate values of \rat\ still lie on the trends
shown.  For NGC 253 and NGC 2146, the disk ratio decreases with the
size of the central bin.  The CO intensity distribution in NGC 253 is
strongly centrally peaked inside 2 kpc.  For NGC 2146, several outer
points are not detected in \ico, so its distribution in the disk
appears flatter than it may be (Figure~\ref{fig.profiles}), and the
change in \rat\ is somewhat biased by this sensitivity limit.  For NGC
3556, NGC 4631 and NGC 6946, the central ratio decreases with the size
of the central bin, and the gradient in \rat\ also decreases.  NGC
3556 and NGC 4631 have flatter CO and \ico\ distributions with peaks
offset from the center by 1.5--2 kpc that are more pronounced in \ico\
emission.  The CO distribution for NGC 6946 peaks more sharply than
the \ico\ at the nucleus.  The central ratio increases with bin size
for IC 342, NGC 3593 and NGC 3627.  Their \ico\ distributions are more
strongly centrally peaked within 1 kpc than the CO.  M82 is unusual in
that \rat\ in the disk increases as the central bin increases from 0.7
to 1.5 kpc radius, but then decreases for a central bin of 2 kpc
radius.

\section{Discussion}

\subsection{Possible Causes of Variations in the CO/\ico\ Intensity
Ratio}

Various physical mechanisms could cause a gradient in the observed
CO/\ico\ intensity ratio.  The most obvious are changes in the
relative CO and \ico\ abundances, and beam-averaged optical depths.  A
gradient in the X-factor, which may depend on the gas density and/or
temperature, could also be responsible.  For the following discussion,
we assume, based on the centrally peaked profiles of integrated
intensity (Figure~\ref{fig.profiles}), that the total beam-averaged
\htwo\ column density increases towards the nucleus of each survey
galaxy.  Also, we assume that \rat\ generally decreases with
galactocentric distance.  (It increases with radius only for NGC 3628,
NGC 5055 and M51, and their individual ratios at large radii are still
low.)

\subsubsection{Abundances}

There is a clear positive gradient in the $^{12}$C/$^{13}$C abundance
ratio in the Milky Way (e.g., Langer \& Penzias 1990, and references 
therein).  It rises linearly from a value near 30 within 4 kpc of the
Galactic center, to $\sim 70$ at 10 kpc, and is most likely due to
stellar processing of matter during the lifetime of the Galaxy; older
populations have enhanced the $^{13}$C in the bulge.  However, this
gradient is in the opposite sense of the gradient in \rat\ seen here
and in the Milky Way.  For example, large-scale maps of the Milky Way
suggest that \rat\ varies from $10\pm 3$ in the central 2 kpc to $6\pm
1$ from 2 to 8.5 kpc (A. Luna, private communication), which is very
similar to the results of this survey.  Though \rat\ appears to be
roughly constant at $\sim 5$ throughout most of the Galaxy (e.g.,
Solomon, Sanders \& Scoville 1979; Polk et~al. 1988; Oka et~al. 1998),
large-scale observations may yield higher values of \rat\ due to the
inclusion of CO emission from very diffuse gas not detectable in \ico\
emission \citep{polk, lee.snell}.  Also, the generally low CO/\ico\
intensity ratios seen in galaxies indicate that at least the CO
emission is optically thick, and therefore \rat\ is a poor tracer of the
$^{12}$C/$^{13}$C abundance ratio.

Another argument against stellar processing affecting \rat\ is that a
correlation should exist between \rat\ and the C$^{18}$O/C$^{17}$O
intensity ratio assuming that the stellar populations in the starburst
nuclei of these galaxies contain relatively few low to intermediate
mass stars (those responsible for much of the \ico\ generation) (Sage,
Henkel \& Mauersberger 1991).  However, such a correlation is not
observed.  In particular, the C$^{18}$O/C$^{17}$O intensity ratio for
M82 of $\sim 8$ is similar to that of NGC 253 and IC 342, whereas the
central CO/\ico\ intensity ratio of M82 is twice those of the other
galaxies.  Casoli, Dupraz, \& Combes (1992) also suggest that the CO
abundance could increase by a factor of two due to an enhanced
production of CO by massive stars.  However, given the high optical
depth of CO, this change will not alter \rat.  Further,
\citet{taniguchi} attribute the high ratios seen in some galactic
nuclei to a {\em deficit} of \ico, not an enhancement of CO.

It has been suggested that the CO isotopic abundances may be altered by
mechanical means as well.  The high central CO/\ico\ intensity ratios
($> 20$) seen in the nuclei of some mergers and ultraluminous
starburst galaxies \citep{casoli,aalto.corat,hm.corat,taniguchi}, are
thought to be due to either transport of less-processed (\ico -poor)
disk gas to the nucleus via an interaction, or destruction of dense
(\ico -rich) clouds by nuclear superwinds.  If gas transport occurs
faster than \ico\ production in the nucleus, any gradient should be
erased.  Models of galactic bars show that abundance gradients are
indeed flattened after roughly $10^9$ yr \citep{friedli}, and that
mass inflow time scales for slightly perturbed spirals are $\lsim
10^9$ yr, and shorter for stronger interactions (Jog \& Das 1992, and
references therein).  That the gradient in \rat\ for the Milky Way
opposes the $^{12}$C/$^{13}$C abundance gradient also implies that
altering abundances seems to have little effect on the CO and \ico\
emission.  Further, NGC 3628 and M51 are two of the few galaxies in
this survey with positive gradients in \rat, yet they are clearly in
interactions.  Therefore, unless a merger injects \ico -poor gas
directly into a nucleus, transport due to bars or interactions seems
unlikely to produce this gradient.  The observed gradient in \rat\
instead favors the superwind argument since \rat\ and the star
formation rate in these galaxies are generally higher in the nucleus
than in the disk.  However, the volume and column densities of the gas
toward these galactic nuclei are typically inferred to be large (e.g.,
Paglione, Jackson, \& Ishizuki 1997).  Therefore, if a superwind
causes the higher CO/\ico\ intensity ratio in the nucleus, it must
force the clouds into many very small, dense clumps.  This possibility
is explored in the following section.

Two chemical processes could be responsible for the gradient in \rat:
isotope-selective photodissociation (ISPD) and chemical fractionation.
ISPD can lower the \ico\ abundance relative to CO in the nucleus since
the rarer isotope is less shielded from destructive radiation.  ISPD
requires a strong, localized source of ultraviolet radiation, and
clumpiness to maximize the surface area of the clouds while
maintaining the high observed column densities.  The nuclei of nearly
all the survey galaxies exhibit high far-infrared (FIR) luminosities
indicative of massive star formation, so sources of dissociating
photons exist.  ISPD would produce CO/C$^{18}$O intensity ratios even
higher than \rat\ since the C$^{18}$O optical depth is lower than that
of \ico.  Nine galaxies have CO, \ico\ and C$^{18}$O emission observed
at roughly the same resolution as this work \citep{casoli.3256,
aalto.corat, sage.c18o}, and there is a strong correlation between
\rat\ and the CO/C$^{18}$O intensity ratio (Figure~\ref{fig.corats}).
However, there is no corresponding positive correlation between \rat\
and \ico/C$^{18}$O.  In addition, \ico/C$^{18}$O stays the same or
even {\em decreases} when viewed with higher resolution in NGC 253 and
M82 \citep{harrison, wild} while it should presumably be highest near
the central ionizing sources.  In fact in M82 at $13''$ resolution,
the highest \ico/C$^{18}$O ratio is found toward the weaker, eastern
millimeter continuum source \citep{wild, carlstrom}.

The \cii/CO intensity ratio should be sensitive to the ultraviolet
field strength and clumpiness of the gas \citep{stacey}, and thus may
also indicate the importance of ISPD.  Figure~\ref{fig.rvsr} shows the
central value of \rat\ for the survey galaxies plotted against this
ratio.  The linear correlation coefficient is less than 0.2,
indicating a very low probability of correlation.

Fractionation in the disk could lower the CO column density relative
to that of \ico\ by factors of a few, but would have little effect on
\rat\ because of the high optical depth of the CO line.  Also, this
mechanism requires lower temperatures ($< 30$ K), and CO optical
depths $< 10$ \citep{keene}.  The CO line strengths indicate
substantial disk column densities ($> 10^{20}\,\cmtwo$), but if the
gas is sufficiently clumpy, fractionation on individual clump surfaces
may be possible.  Whether the cloud  temperatures in the disk are low
enough to support large-scale fractionation is more difficult to
determine.  The temperatures of the central regions of all the survey
galaxies are inferred to be high from multi-line observations of CO
\citep{wall.kpc, wild}, \cii\ spectra \citep{stacey}, and IRAS colors
\citep{young}, but little data for galactic disks exist.  Studies of
NGC 6946 and M51 indicate that higher gas temperatures in the disk are
limited to the spiral arms and various hot spots \citep{engargiola,
fitt, madden, tuffs}.  Little CO emission is seen outside of these
warmer regions, where the gas might be cool enough to support
fractionation.

To summarize, we eliminate nearly all arguments in favor of abundance
variations as the cause of a gradient in \rat.  ISPD and CO
enhancements due to gas transport or stellar nucleosynthesis are
unlikely to be responsible for the observed gradients in \rat\ due to
the high optical depth of CO, and the lack of expected gradients or
correlations with other line ratios.  Also the gradient in \rat\ in
the Milky Way is in the opposite sense of the $^{12}$C/$^{13}$C
abundance gradient.  It is difficult to test whether fractionation in
the disk is important, though most of the CO emission in galactic
disks is limited to relatively warm regions where it would not occur.
Nuclear superwinds could effectively lower the beam-averaged \ico\
column density by creating very small clouds, which would result in
higher values of \rat\ in starburst nuclei.  This prediction is tested
in the following section. 

\subsubsection{Cloud Properties}

The CO/\ico\ intensity ratio should be a straightforward measure of
the \ico\ optical depth.  For example in LTE, given optically thick CO
emission and optically thin \ico\ emission, and that their emission
comes from the same volume of gas, 

\begin{equation}
I({\rm CO})/I(\ico) \approx 1/\tau(\ico).
\end{equation}

\noindent
In this case, the mean optical depth of the gas in these galaxies
generally decreases toward the nucleus.  In LTE, $\tau \propto
(N/\Delta v)/T_k^2$, for high $T_k$.  Therefore, to produce the
observed gradient in \rat, the clouds in galactic nuclei are probably
warmer, have lower column densities, and/or higher internal velocity
dispersions.

To determine which cloud parameters dominate the variation in \rat,
and to include volume density as another important cloud parameter, we
perform single-component model calculations of non-LTE CO excitation.
We assume that the emission originates in unresolved, homogeneous,
spherical clouds, and include a photon escape probability function to
account for the radiative excitation of optically thick lines
\citep{rt}.  The emission from the first 11 levels of CO is modeled.
We use the collision rates of \citet{co.coll}.  The excitation of
\ico\ is assumed to be identical to that of CO \citep{wlg}, and the
\ico\ line temperature is estimated by decreasing the column density
of CO by the [CO]/[\ico] abundance ratio.  Values of
[CO]/[\ico]=30--80 are considered.

Figure~\ref{fig.model} shows the expected CO/\ico\ intensity ratio
from a cloud as a function of \htwo\ volume density (\nhh), CO column
density per velocity interval ($N/\Delta v$), and kinetic temperature
($T_k$), given [CO]/[\ico]=60.  As expected, the ratio is a strong
function of cloud column density and temperature, especially if the CO
emission is thermalized.  For example, a change in $T_k$ from 10 to 30
K results in roughly a factor 3 variation in \rat.  For low densities,
the optical depth of \ico\ is non-neglible, and the dependence on
temperature diminishes (though a higher optical depth of the \ico\
line does not necessarily imply \rat\ $\sim 1$ when LTE does not
apply).  Its dependence on density is weaker, especially for $T_k <
50$ K, and reflects the difference between subthermally excited and
thermalized (near LTE) CO emission.  Below \nhh $\sim 10^5$ \percc,
\rat\ may vary by factors of a few with density, depending on $T_k$
and $N/\Delta v$.

To test how variations in $N/\Delta v$ may affect the gradient in
\rat, we estimate the ratio of the column density within and outside
of a 2 kpc radius in each galaxy.  The column density per velocity
interval for individual clouds, $N_c/\Delta v_c$ would have to
decrease toward the nucleus to cause an increase in \rat.  We also
assume a proportionality between the CO and \ico\ intensities and the
beam-averaged column density, $N$, such that

\begin{equation}
\frac{(N_c/\Delta v_c)_{R<2 {\rm kpc}}}{(N_c/\Delta v_c)_{R>2 {\rm kpc}}} =
	\frac{(X I/\Delta v)_{R<2 {\rm kpc}}}{(X I/\Delta v)_{R>2 {\rm kpc}}}
	\Bigl(\frac{\phi_{R>2 {\rm kpc}}}{\phi_{R<2 {\rm kpc}}}\Bigr),
\end{equation}

\noindent
where $\phi$ is the beam filling factor, and we assume constant CO and
\ico\ abundances.  The velocity width is estimated from the full line
width.  $I/\Delta v$ rises toward the nuclei of each survey galaxy by
factors of 2--30 using CO, and 1.3--7 using \ico.  For individual
clouds, $N_c/\Delta v_c$ could still decrease towards the nucleus if
the beam filling factor rises faster than $X I/\Delta v$.  In other
words, if crowding in the nucleus is significant, then a higher value
of \rat\ is still possible if $N_c/\Delta v_c$ of individual clouds
there is lower than in the disk.  Such increases in $\phi$ appear
reasonable given the increase in line intensities and their likely
dependence on varying excitation conditions from disk to nucleus.
Therefore, we cannot eliminate the possibility of small nuclear
clouds, such as would be produced by starburst winds, though the
amount of variation in $N_c/\Delta v_c$ depends mostly on the internal
turbulence of the clouds and the change in excitation conditions, that
is, the variation in volume and column density and kinetic
temperature.

An increase in the gas volume density toward the nucleus can
contribute to the gradient in \rat.  For certain conditions ($T_k >$
10 K, \nhh $\sim 10^3$--$10^5$ \percc), an increase in density of an
order of magnitude can double \rat.  The centers of galaxies are known
to have enhanced gas densities \citep{wall.kpc, hb.sd, hcn32}.
Figure~\ref{fig.rvsr} shows the central value of \rat\ for the survey
galaxies plotted against the (HCN \jth)/(HCN \jon) intensity ratio,
which is sensitive to the gas density \citep{hcn32}.  A positive trend
is apparent, though there is no significant correlation (the linear
correlation coefficient is 0.2).

The correlation between \rat\ and IRAS color temperature (the ratio
of global fluxes at 60 and 100$\mu$m) is well established
\citep{young.sanders, aalto.corat, sage.isbell, taniguchi}.  There is
a good correlation between \rat\ and color temperature for the
galaxies presented here as well.  According to the homogeneous cloud
model, \rat\ varies with the gas kinetic temperature also, especially
for \nhh\ $> 10^3$ \percc.  Generally \rat\ $\propto T_k^{1.4}$ for
high densities, and \rat\ $\propto T_k^{0.3}$ for \nhh\ $< 10^3$ 
\percc.  Such a strong dependence on temperature, especially in
combination with the expected increase in density towards the nucleus,
suggests that most if not all of the variation seen in \rat\ is due to
the high gas temperatures in these (mostly starburst) galactic
nuclei.  Figure~\ref{fig.rvsr} shows the central value of \rat\ for
the survey galaxies plotted against the (CO \jth)/(CO \jon) intensity
ratio, which is generally sensitive to $T_k$.  A linear correlation
coefficient of 0.5 indicates a significant correlation between these
ratios.

A two-component model can also explain a rise in \rat\ at the nucleus
without gradients in the bulk physical conditions of the clouds.  We
test a model where the cloud properties do not change, but warm and
diffuse gas ($\nhh = 100$ \percc, $T_k = 30$ K) is added to the
nucleus.  Given the observed CO/\ico\ ratios in galactic disks, a
diffuse component with twice the beam filling factor of the dense
component provides enhancements in \rat\ as high as 90\%\ for very
low temperatures ($T_k < 20$ K).  Enhancements of 10--50\%, for $T_k
= 20$--75 K and $\nhh \sim 10^2$--$10^4$\percc, are more likely and
still match the observed gradients in \rat.  Decrements in \rat\ of
up to 20--40\%\ can also be produced by including the diffuse
component, but require that the dense component be very dense and
warm ($\nhh > 10^4$ \percc, $T_k > 50$ K), which are unlikely
conditions in galactic disks.

To summarize, the observed gradient in \rat\ can be successfully
explained by a gradient in the gas kinetic temperature.  Increased
volume densities in the nucleus may also contribute to raising \rat\
there.  The column densities of individual clouds may decrease enough
to cause the observed gradient in \rat, though this variation may be
secondary to changes in temperature.  This result is expected since in
the LTE limit, the optical depth of \ico, which is inversely
proportional to \rat, depends more strongly on $T_k$.  Therefore a
modest rise in $T_k$ towards the nucleus (e.g., from 10 K to 30 K, see
Figure~\ref{fig.model}) can easily account for the gradient in \rat.
Finally, a two-component model with warm and diffuse gas in the
nucleus can reproduce the observed enhancements in \rat\ for a
reasonable range of temperatures and densities.  However, the observed
decrements in \rat\ are more difficult to explain with this model.

\subsubsection{X-Factor}

A positive gradient in the X-factor has been suggested for the Galaxy
\citep{grad.xco, sod.gc, dahmen}.  This result would follow naturally
from the gradient in \rat\ found in this survey and the Milky Way.
That is, the \htwo\ mass estimated from a direct proportionality with
CO intensity would overestimate the true mass in galactic centers,
which is presumably traced less ambiguously by the more optically thin
\ico\ emission.  Despite the probable Galactic gradient in the
X-factor, the apparently strong dependence of \rat\ on cloud
properties (and possibly chemical abundances) indicates that many
things can contribute to varying the proportionality between $I_{CO}$
and \htwo\ column density.  Using the homogeneous cloud model from the
previous section, as expected, \rat\ increases as $X$ decreases given
a constant CO abundance. Therefore, we suggest that a variable
X-factor results from the same processes that affect \rat.
Determining unambiguous variations in $X$ requires a better
understanding of the many physical processes that can affect \rat.

We estimate the \htwo\ column densities inside and outside a 2 kpc
radius in the survey galaxies using various means: from the CO
intensity and the standard, constant X-factor of
1.6\ee{20}\,\cmtwo\,(\kkms)$^{-1}$; from the \ico\ intensity,
assuming optically thin emission and LTE; and from the homogeneous
cloud model, which includes non-LTE excitation and radiative trapping.

The molecular masses inside a 2 kpc radius, derived from the standard
X-factor, are listed in Table~\ref{tab.cols}.  For most of the
galaxies, the mass derived from CO is less than the dynamical mass
inside 2 kpc.  However for IC 342, M82 and NGC 6946, it is comparable
to the dynamical mass.  Unlike M82, IC 342 and NGC 6946 have typical
CO/\ico\ intensity ratios.

The \htwo\ column density can be estimated from the \ico\ intensity
assuming optically thin emission and LTE using

\begin{equation}
N_\htwo\ = 4.2\ee{19} T_k e^{2.645/T_k} I(\ico),
\label{eq.x13}
\end{equation}

\noindent
where the \ico\ abundance is 8\ee{-5}/60 \citep{co.abund}.  Masses
derived from equation~\ref{eq.x13} are listed in Table~\ref{tab.cols}
for $T_k=10$ K and 50 K.  For most of the survey galaxies, these
masses are similar to those derived from the standard X-factor and
CO, which implies that the kinetic temperature is between 10 and 50
K.  In fact, the gas kinetic temperature can be estimated by setting
the column densities derived from both CO and \ico.  The temperatures
inside 2 kpc (Table~\ref{tab.cols}) are 20--60 K, and only somewhat
lower in the disk.  The higher temperatures are quite high compared to
those derived from radiative transfer analyses \citep{wall.kpc}, and
seem unlikely over such large areas, especially in galactic disks.
That this temperature seems high indicates that the masses derived
from \ico\ are generally lower than those derived from CO.

We estimate column densities with the homogeneous cloud model assuming
\nhh\ = 300 \percc\ and $T_k = 10$ K in the disk, and \nhh\ = $10^4$
\percc\ and $T_k = 30$--50 K within a 2 kpc radius.  We use the
CO/\ico\ intensity ratios given in Table~\ref{tab.rat}.  Assuming LTE
and optically thin \ico\ emission, the kinetic temperature is
overestimated on average by a factor of 3--4 in the disk, and
underestimated by 2.3--3 times in the nucleus.  The average conversion
factor in the disk is (1--3)\ee{20}\,\cmtwo\,(\kkms)$^{-1}$, which is
consistent with the value in the disk of the Milky Way.  For the
nuclei we find $X$ = (0.4--1.4)\ee{20}\,\cmtwo\,(\kkms)$^{-1}$, which
is factors of 2--5 times lower than in the disk, and 1.2--3.3 times
lower than the standard Galactic value.  Using the standard Galactic
X-factor in galactic nuclei could therefore overestimate the true mass
by factors of a few.  Note that the model indicates that the X-factor
is always smaller in galactic nuclei, regardless of the gradient in
\rat.

\section{Conclusions}

We observed the CO and \ico\ \jon\ emission along the major axes of 17
nearby galaxies.  On average, the CO/\ico\ intensity ratio, \rat,
inside a 2 kpc radius is roughly 30\%\ higher than in the disk.  This
ratio is sensitive to variations in temperature and column density, as
well as fractionation and isotope-selective photodissociation.  Winds
and gas inflow caused by mergers have also been suggested to vary
\rat, along with abundance variations due to stellar processing.  We
eliminate most mechanisms except for fractionation and variations in
kinetic temperature, cloud column density and \htwo\ volume density.
The gradient is most likely caused by the higher temperatures (and
perhaps densities) typical of the central regions of starburst
galaxies.  Small nuclear clouds, perhaps caused by starburst
superwinds, may also contribute to the variation in \rat.  We
estimate that the X-factor ($X=N_\htwo/I_{CO}$) decreases toward
galactic nuclei, as seen in the Milky Way, and its variation results
from the same physical processes that affect \rat.  A modest increase
in global gas temperatures easily accounts for the observed variation
in \rat, as well as a decrease in $X$ of factors of 2--5, from disk
to nucleus.

\acknowledgements

The FCRAO is operated through National Science Foundation grant
AST 88-15406 and with the permission of the Metropolitan District
Commission.  This work was also supported in part by grant
211290-5-25875E from the Consejo Nacional de Ciencia y
Tecnolog\'{\i}a.  The authors are grateful to M. Skrutski and the
UMass Society of Physics Students for their support and help with
recruiting interested young scientists.  We also thank M. Belniak,
N. Bouch, H. Cameron, N. Cartel, N. Durham, M. Gruol, S. Litchfield,
J. Perry, S. Stephan, M. Thorn and R. White for their assistance with
the observations.  M. Richard and M. Goldstein thank the Five College
Astronomy Department and the FCRAO for summer internships.
T. Paglione gratefully acknowledges the generous hospitality of
A. Lovell and D. Smith.

\clearpage

\begin{center}
\begin{deluxetable}{lccccccc}
\tablewidth{0pt}
\tablecaption{Properties of Survey Galaxies and Maps.
\label{tab.sources}
}
\tablehead{
Galaxy & R.A. (2000)& Decl. (2000) & \vlsr$^a$ & $D$ &
 Map P.A. & $d^b$ & Map Size$^c$ \\
 & h m s & $\degr$ $'$ $''$ & (\kms) & (Mpc) & (degrees) & (kpc) & (kpc)
}
\startdata
NGC 253  & 00 47 35.1 & $-$25 17 20\phm{$-$} & 230 & \phn3 & 54 & 0.7
& \phn9.3\\ 
NGC 1068 & 02 42 40.7 & $-$00 00 48\phm{$-$} & 1150\phn & \phn\phd14.4
& 90 & 3.2 & 20.1\\ 
IC 342   & 03 46 49.7 & 68 05 45 & \phn 30 & \phn4 & \phn 0 & 0.9 &
15.9\\ 
UGC 2855 & 03 48 22.6 & 70 07 57 & 1200\phn & 20 & 92 & 4.5 & 30.1\\ 
NGC 2146 & 06 18 37.6 & 78 21 19 & 920 & 14 & 128\phn & 3.1 & 13.5\\ 
M82      & 09 55 54.0 & 69 40 57 & 250 & \phn\phn\phn\phd3.25 & 65 &
\phn0.72 & \phn8.4\\ 
NGC 3079 & 10 01 58.2 & 55 40 42 & 1150\phn & 20 & 164\phn & 4.5 &
27.9\\ 
NGC 3184 & 10 18 17.2 & 41 25 26 & 590 & 13 & 135\phn & 2.9 & 23.7\\ 
NGC 3556 & 11 11 31.8 & 55 40 14 & 700 & 10 & 80 & 2.2 & 15.0\\ 
NGC 3593 & 11 14 36.0 & 12 49 06 & 600 & \phn9& 90 & 2.0 & 12.6\\ 
NGC 3627 & 11 20 14.4 & 12 59 42 & 740 & \phd\phn\phn6.7 & \phn0 & 1.5
& 12.2\\
NGC 3628 & 11 20 16.2 & 13 35 22 & 850 & \phd\phn\phn6.7 & 103\phn &
1.5 & 15.1\\
NGC 4527 & 12 34 08.8 & 02 39 13 & 1770\phn & 15 & 67 & 3.3 & 20.9\\
NGC 4631 & 12 42 07.6 & 32 32 28 & 630 & \phn9& 88 & 2.0 & 13.5\\
NGC 5055 & 13 15 49.2 & 42 02 06 & 500 & \phn7 & 105\phn & 1.6 &
15.8\\ 
M51      & 13 29 53.2 & 47 11 48 & 450 & \phd\phn\phn9.6 & \phn 0 &
2.1 & 21.6\\
NGC 6946 & 20 34 51.8 & 60 09 15 & \phn 50 & 10 & 45 & 2.2 & 22.5\\
\enddata
\tablenotetext{a}{Central position}
\tablenotetext{b}{Linear extent of 46$''$ at distance $D$}
\tablenotetext{c}{Full extent of map along major axis, sampled every
$22\farcs14$ with a 46$''$ beam (FWHM).}
\end{deluxetable}
\end{center}

\begin{center}
\begin{deluxetable}
{lr@{ $\pm$ }l@{ $\pm$ }lr@{ $\pm$ }lr@{ $\pm$ }lr@{ $\pm$ }lr@{ $\pm$ }l}
\tablewidth{0pt}
\tablecaption{Variation of CO/\ico\ Intensity Ratio.
\label{tab.rat}
}
\tablehead{
Galaxy & \multicolumn{9}{c}{$I$(CO)/$I$(\ico)$^a$}
 & \multicolumn{2}{c}{Ratio$^b$} \\
\cline{2-10}
 & \multicolumn{3}{c}{$R< 23''$} & \multicolumn{2}{c}{$R> 23''$}
 & \multicolumn{2}{c}{$R< 2$ kpc} & \multicolumn{2}{c}{$R> 2$ kpc}
 & \multicolumn{2}{c}{}
}
\startdata
NGC 253  & 11.9 & 0.5 & 3.4 &  8.8 & 0.5 & 12.4 & 0.5 &  6.9 & 0.8 &
1.8 & 0.2 \\
NGC 1068 & 10.7 & 0.6 & 1.9 & 11.7 & 1.2 & 10.4 & 0.4 & 11.6 & 1.1 &
0.9 & 0.1 \\
IC 342   &  8.4 & 0.5 & 1.6 &  7.1 & 0.6 & 10.8 & 0.8 &  6.4 & 0.6 &
1.7 & 0.2 \\
UGC 2855 & 11.7 & 1.7 & 2.3 &  9.4 & 1.7 & 11.7 & 1.7 &  9.4 & 1.7 &
1.2 & 0.3 \\
NGC 2146 & 14.9 & 1.3 & 8.6 & 28.1 & 6.9 & 15.5 & 1.1 & 22.8 & 4.5 &
0.7 & 0.1 \\
M82      & 27.3 & 1.0 & 7.8 & 12.9 & 0.5 & 20.9 & 0.7 & 12.1 & 6.9 &
1.7 & 1.0 \\
NGC 3079 & 15.3 & 1.8 & 5.1 & 10.1 & 2.1 & 15.3 & 1.8 & 10.1 & 2.1 &
1.5 & 0.4 \\
NGC 3184 &  4.9 & 1.2 & 2.3 &  4.4 & 0.8 &  5.6 & 1.1 &  4.0 & 0.6 &
1.4 & 0.3 \\
NGC 3556 &  8.7 & 1.4 & 5.6 &  6.8 & 0.6 &  7.0 & 0.8 &  6.9 & 0.7 &
1.0 & 0.2 \\
NGC 3593 & 10.6 & 2.2 & 10  & 11.4 & 3.2 & 16.8 & 4.6 &  5.7 & 2.3 &
2.9 & 1.4 \\
NGC 3627 & 11.2 & 1.2 & 5.7 & 17.4 & 2.1 & 15.5 & 1.6 & 17.8 & 4.1 &
0.9 & 0.2 \\
NGC 3628 &  9.2 & 0.4 & 2.3 & 10.3 & 0.6 &  9.7 & 0.4 & 11.2 & 1.1 &
0.9 & 0.1 \\
NGC 4527 &  6.1 & 0.5 & 1.4 &  6.9 & 0.8 &  6.7 & 0.9 &  6.7 & 0.6 &
1.0 & 0.2 \\
NGC 4631 & 16.3 & 2.4 & 9.9 &  9.1 & 0.7 & 13.1 & 1.3 &  8.4 & 0.9 &
1.6 & 0.2 \\
NGC 5055 &  5.7 & 0.6 & 3.1 &  8.2 & 0.7 &  7.1 & 0.6 &  9.2 & 1.1 &
0.8 & 0.1 \\
M51      &  5.4 & 0.4 & 1.0 &  7.5 & 0.6 &  6.4 & 0.4 &  8.2 & 0.9 &
0.8 & 0.1 \\
NGC 6946 & 17.0 & 1.4 & 6.6 & 10.4 & 1.1 & 12.7 & 1.0 &  9.8 & 1.4 &
1.3 & 0.2 \\[10pt]
Average  & 11.5 & 0.3 & 1.4 & 10.6 & 0.5 & 11.6 & 0.4 &  9.8 & 0.6 &
1.3 & 0.1 \\
\enddata
\tablenotetext{a}{From weighted averages of data within or beyond noted
radii.  For the central position, the estimated systematic uncertainty
is listed after the statistical uncertainty.}
\tablenotetext{b}{Intensity ratio inside 2 kpc divided by that outside
of 2 kpc.  Uncertainty is based on the statistical uncertainty.}
\end{deluxetable}
\end{center}

\begin{center}
\begin{deluxetable}{lccc@{\ }c@{\ }c@{\ }c@{\ }c}
\tablewidth{0pt}
\tablecaption{Modeling Results$^a$.
\label{tab.cols}
}
\tablehead{
Galaxy & $M_\htwo$(CO) & \multicolumn{2}{c}{$M_\htwo$(\ico)} &
\multicolumn{2}{c}{$T_k$ (K)} & \multicolumn{2}{c}{$X$} \\
\cline{3-4} \cline{5-6} \cline{7-8}
& & $T_k=10$ K & $T_k=50$ K &
$R<2$ kpc & $R>2$ kpc & $R<2$ kpc & $R>2$ kpc
}
\startdata

NGC 253  & 2.40 (0.04) & 0.66 (0.03) & 2.7 (0.1) & 45 (2) & 23 (3) &0.5$_{-0.2}^{+0.4}$ & 2.5$_{-0.7}^{+1.5}$ \\
NGC 1068 & 2.16 (0.32) & 0.75 (0.10) & 3.0 (0.4) & 37 (2) & 42 (4) &0.7$_{-0.2}^{+0.2}$ & 1.1$_{-0.3}^{+0.4}$ \\
IC 342   & 0.99 (0.16) & 0.32 (0.03) & 1.3 (0.1) & 39 (3) & 22 (2) &0.6$_{-0.2}^{+0.3}$ & 2.8$_{-0.7}^{+0.9}$ \\
UGC 2855 & 0.87 (0.03) & 0.25 (0.04) & 1.0 (0.1) & 42 (7) & 33 (6) &0.6$_{-0.2}^{+0.3}$ & 1.6$_{-0.4}^{+0.5}$ \\
NGC 2146 & 1.47 (0.29) & 0.34 (0.09) & 1.4 (0.4) & 56 (4) & 84 (17)&0.4$_{-0.3}^{+1.0}$ & 0.4$_{-0.2}^{+1.1}$ \\
M82      & 2.83 (0.01) & 0.46 (0.01) & 1.9 (0.1) & 77 (3) & 44 (26)&0.3$_{-0.1}^{+0.2}$ & 1.1$_{-0.4}^{+0.7}$ \\
NGC 3079 & 1.87 (0.02) & 0.42 (0.05) & 1.7 (0.2) & 56 (7) & 36 (8) &0.4$_{-0.2}^{+0.4}$ & 1.4$_{-0.5}^{+1.1}$ \\
NGC 3184 & 0.16 (0.01) & 0.09 (0.02) & 0.4 (0.1) & 19 (4) & 13 (2) &1.5$_{-0.7}^{+2.9}$ & 4.9$_{-1.8}^{+8.5}$ \\
NGC 3556 & 0.34 (0.01) & 0.17 (0.02) & 0.7 (0.1) & 24 (3) & 23 (3) &1.1$_{-0.6}^{+5.0}$ & 2.5$_{-1.3}^{+8.7}$ \\
NGC 3593 & 0.50 (0.04) & 0.10 (0.03) & 0.4 (0.1) & 61 (17)& 19 (9) &$>0.1$              & $>1.2$ \\
NGC 3627 & 0.81 (0.01) & 0.18 (0.02) & 0.7 (0.1) & 57 (6) & 65 (15)&0.4$_{-0.3}^{+0.7}$ & 0.5$_{-0.2}^{+1.2}$ \\
NGC 3628 & 1.47 (0.02) & 0.52 (0.02) & 2.1 (0.1) & 34 (2) & 40 (4) &0.7$_{-0.3}^{+0.5}$ & 1.2$_{-0.4}^{+0.6}$ \\
NGC 4527 & 1.03 (0.10) & 0.49 (0.12) & 2.0 (0.5) & 23 (3) & 23 (2) &1.2$_{-0.4}^{+0.8}$ & 2.6$_{-0.7}^{+1.2}$ \\
NGC 4631 & 0.83 (0.02) & 0.22 (0.02) & 0.9 (0.1) & 47 (5) & 29 (3) &0.5$_{-0.3}^{+2.0}$ & 1.8$_{-1.0}^{+7.1}$ \\
NGC 5055 & 0.57 (0.02) & 0.27 (0.02) & 1.1 (0.1) & 25 (2) & 32 (4) &1.1$_{-0.5}^{+2.5}$ & 1.6$_{-0.8}^{+3.0}$ \\
M51      & 1.16 (0.02) & 0.62 (0.04) & 2.5 (0.2) & 22 (2) & 29 (4) &1.3$_{-0.4}^{+0.7}$ & 1.9$_{-0.4}^{+0.7}$ \\
NGC 6946 & 1.79 (0.03) & 0.48 (0.04) & 2.0 (0.1) & 46 (4) & 35 (5) &0.5$_{-0.3}^{+0.5}$ & 1.5$_{-0.6}^{+1.5}$ \\
\enddata
\tablenotetext{a}{Uncertainty given in parentheses.}
\tablenotetext{\ }{Column 1: Galaxy}
\tablenotetext{\ }{Column 2: Mass inside 2 kpc from standard
conversion factor and $I$(CO) in $10^9$ \msun}
\tablenotetext{\ }{Columns 3 and 4: Mass inside 2 kpc from \ico\
(eq.~\ref{eq.x13}) for given $T_k$ in $10^9$ \msun}
\tablenotetext{\ }{Columns 5 and 6: $T_k$ from standard X-factor and
column density estimated from \ico\ (eq.~\ref{eq.x13})}
\tablenotetext{\ }{Columns 7 and 8: X-factor from non-LTE calculations
and \rat\ (Table~\ref{tab.rat}) in $10^{20}$ \cmtwo (\kkms)$^{-1}$}
\end{deluxetable}
\end{center}

\clearpage

\figcaption[f1*.ps]
{Spectra of CO (heavy lines) and \ico\ (light lines) emission for the
survey galaxies.  Except for IC 342, all \ico\ spectra are smoothed to
27.2 \kms\ resolution for display.  The temperature scale is for the
\ico\ line.  The CO line is divided by 10 for display.
\label{fig.spectra}}

\figcaption[f2*.ps]
{The CO and \ico\ integrated intensities ($\int \tas dv$), and their
ratios, versus position along the major axis.  Upper and lower limits
(3$\sigma$) are denoted with downward- and upward-pointing triangles,
respectively.  Error bars are 1$\sigma$ statistical estimates based on
the r.m.s. noise, line width and channel width.
\label{fig.profiles}}

\figcaption[f3.ps]
{The CO/\ico\ intensity ratio as a function of position.  The data are
folded about the nucleus and binned to make the uncertainty in \rat\
uniform along the major axis.  Vertical error bars indicate the
statistical uncertainties of the data (see text).  Horizontal bars
indicate the bin sizes.
\label{fig.rbin}}

\figcaption[f4.ps]
{Comparison of the ratio of CO and \ico\ integrated intensities inside
and outside of selected radii: 23$''$ (beam HWHM), 1, 1.5 and 2 kpc.
The lines indicate slopes of 2/3, 1, 3/2, 2 and 4 (dash-triple-dot,
solid, dashed, dash-dot and dotted, respectively).
\label{fig.nucdisk}}

\figcaption[f5.ps]
{Same as Figure~\ref{fig.nucdisk} indicating the variation in the
gradient of \rat\ with the size of the central region.  The ratios
for M82 are divided by 10 for clarity.  The filled symbol denotes the
ratio for the central point without binning.  The other connected
points are for central regions of 1, 1.5 and 2 kpc radii, in sequence.
\label{fig.drat}}

\figcaption[f6.ps]
{Central CO/\ico\ intensity ratio plotted against the central
CO/C$^{18}$O and \ico/C$^{18}$O intensity ratios.  All data are at
similar resolution, $\sim 45''$.  References - NGC 253, IC 342, M82:
this work, \citet{sage.c18o}; NGC 1808, NGC 2146, NGC 4826, Circinus,
NGC 7552: this work, \citet{aalto.corat}; NGC 3256:
\citet{casoli.3256}.
\label{fig.corats}}

\figcaption[f7.ps]
{Central CO/\ico\ intensity ratio plotted against \cii/CO
\citep{stacey, carral}, HCN \jth/\jon\ \citep{hcn32}, and CO
\jth/\jon\ \citep{mau.co32, n1068.co32, wielebinski}.  Error bars
denote systematic uncertainties.
\label{fig.rvsr}}

\figcaption[f8.ps]
{Expected CO/\ico\ intensity ratios as functions of kinetic
temperature, CO column density per velocity interval, and \htwo\
volume density.
\label{fig.model}}

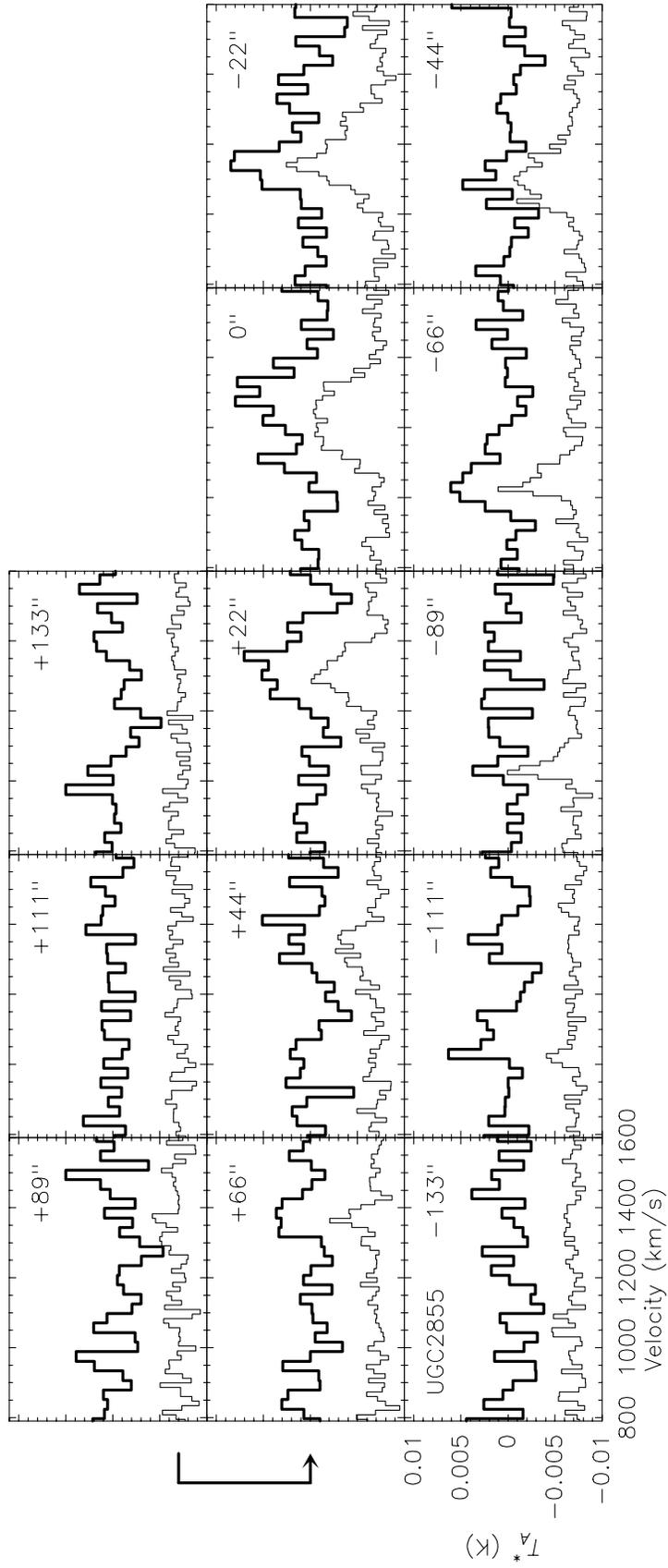
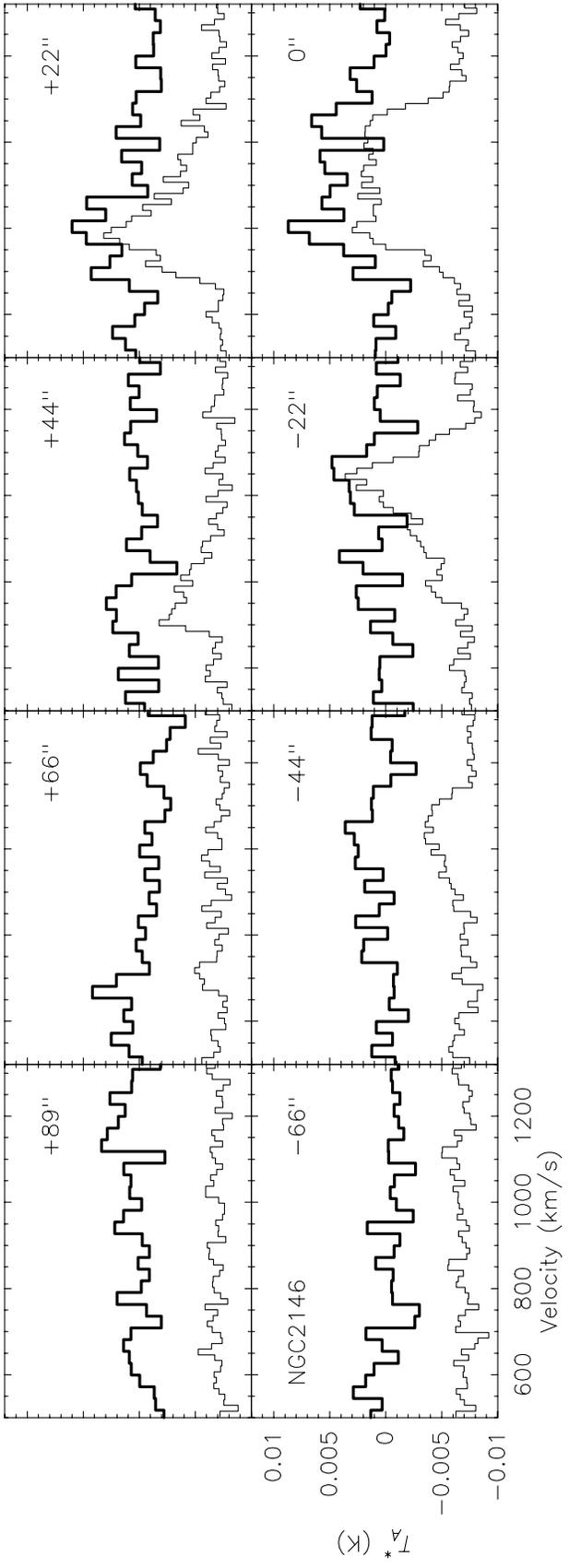
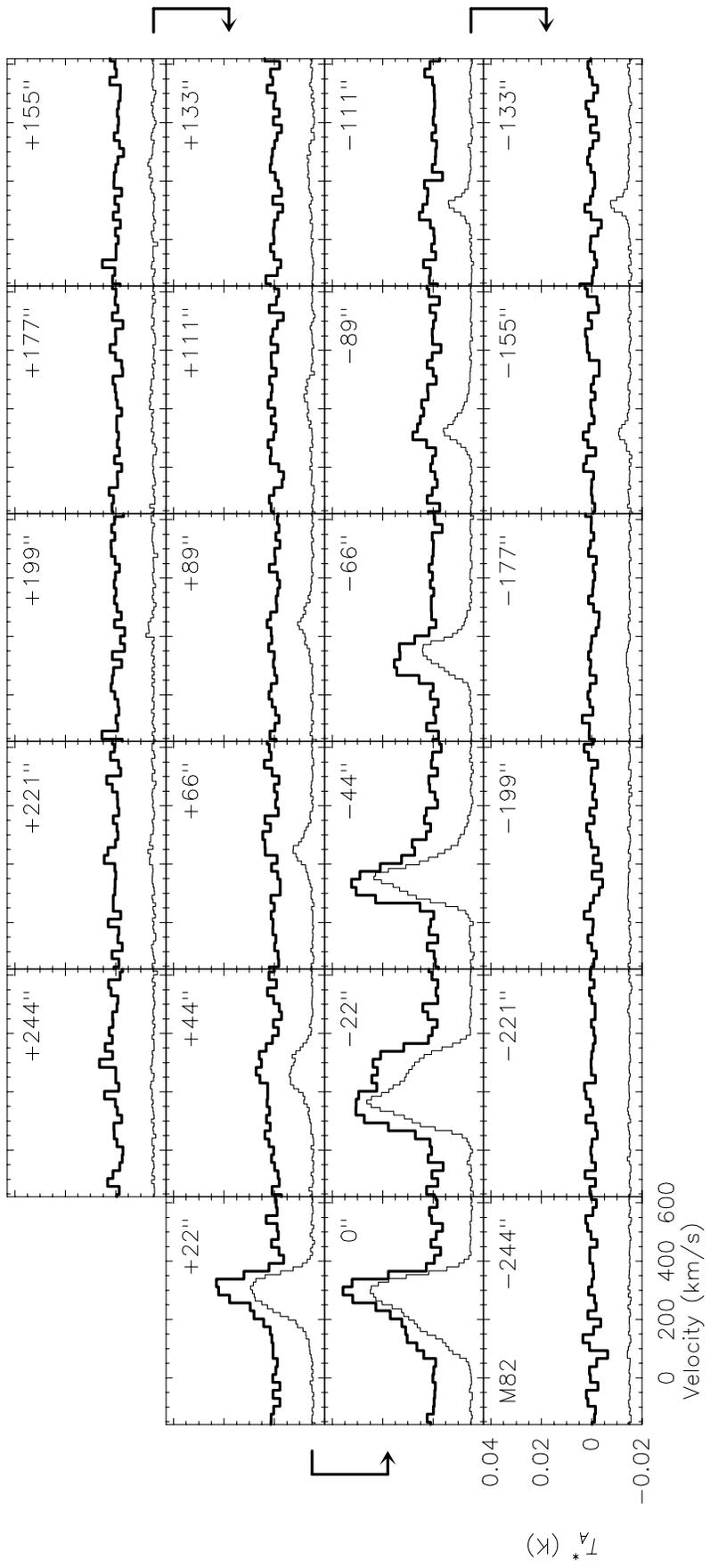
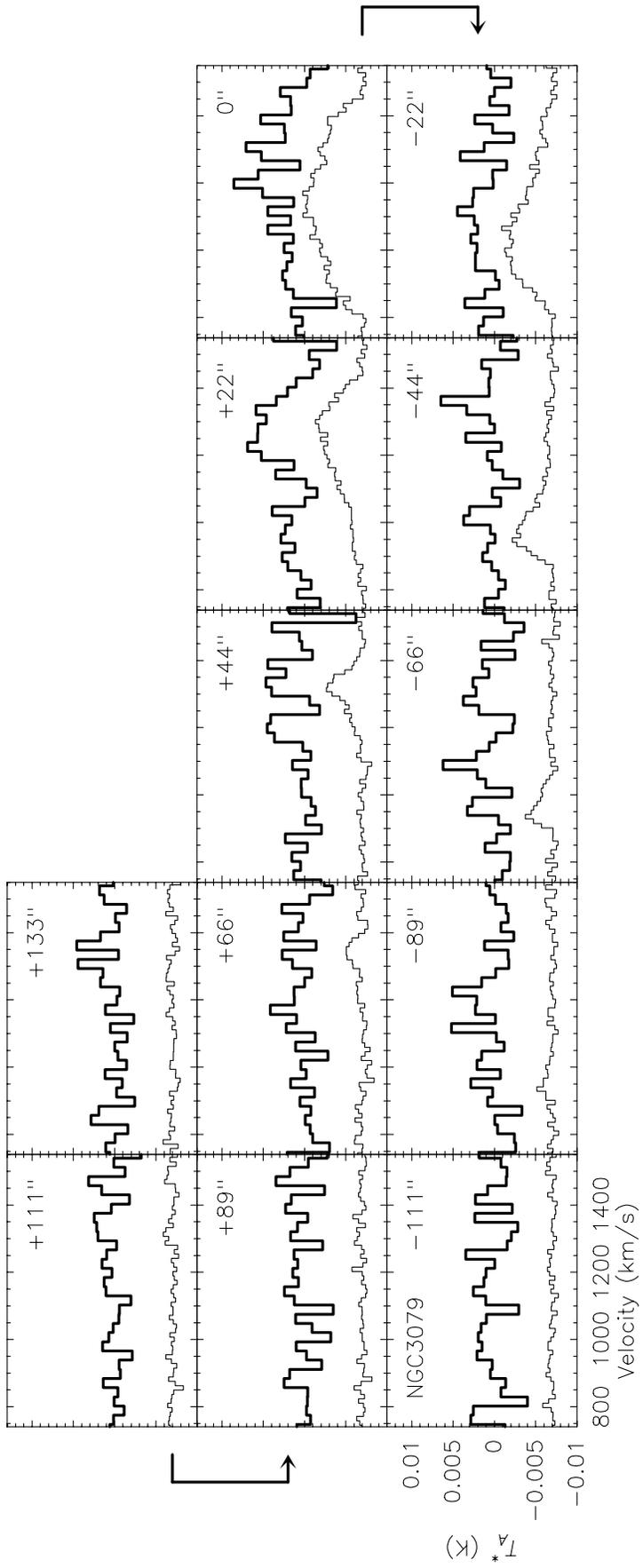
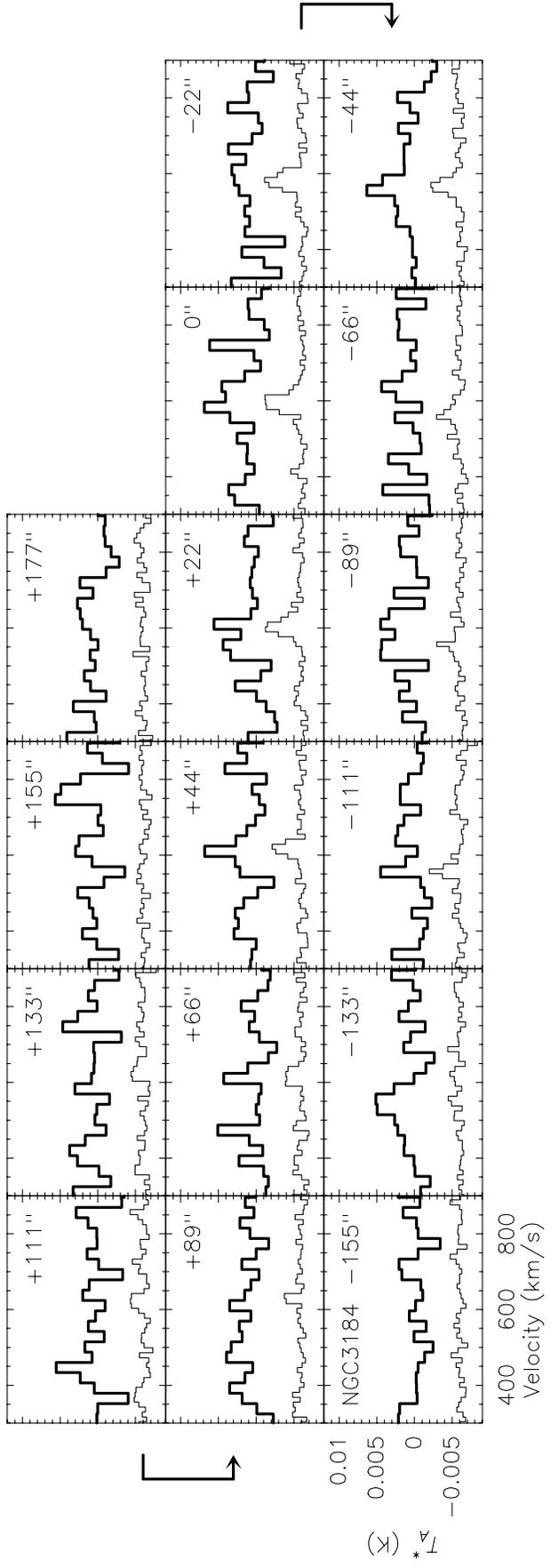
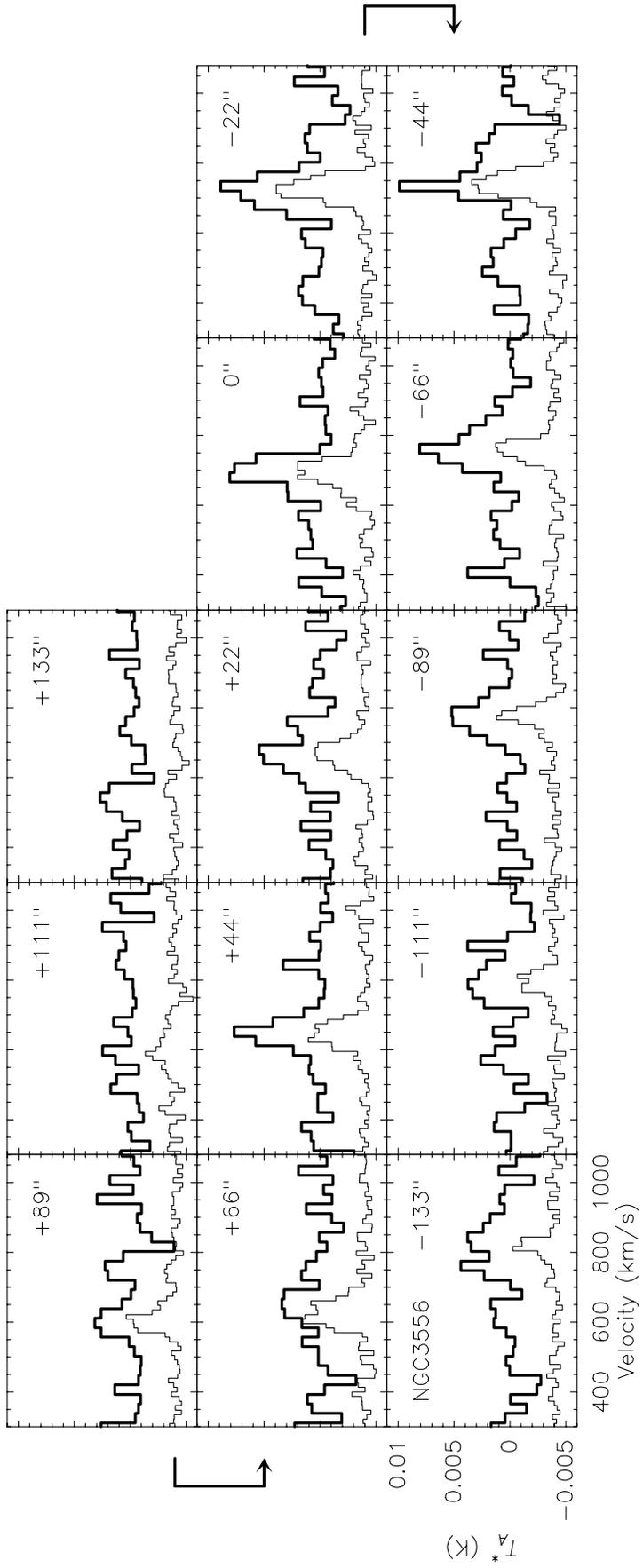
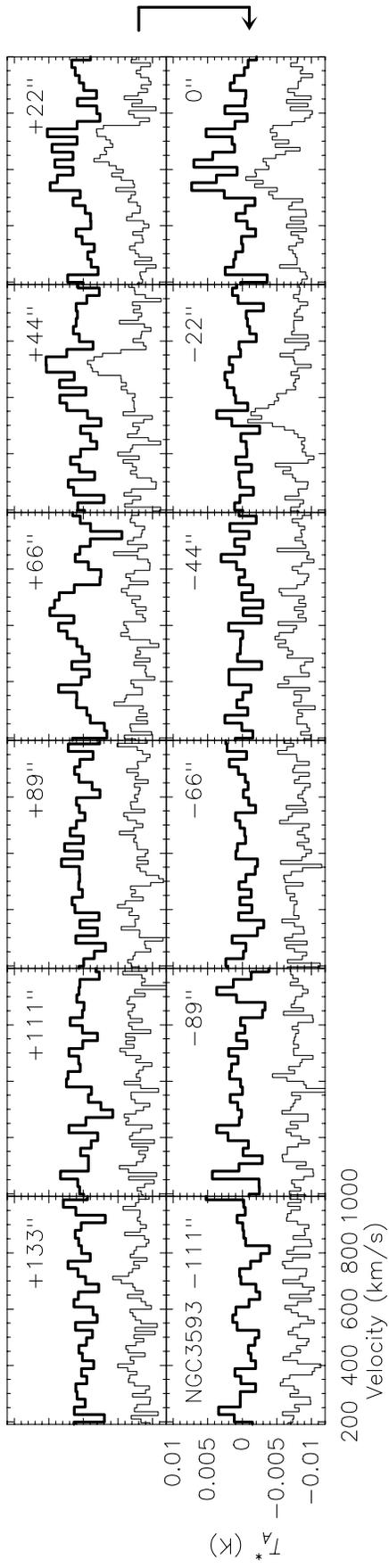
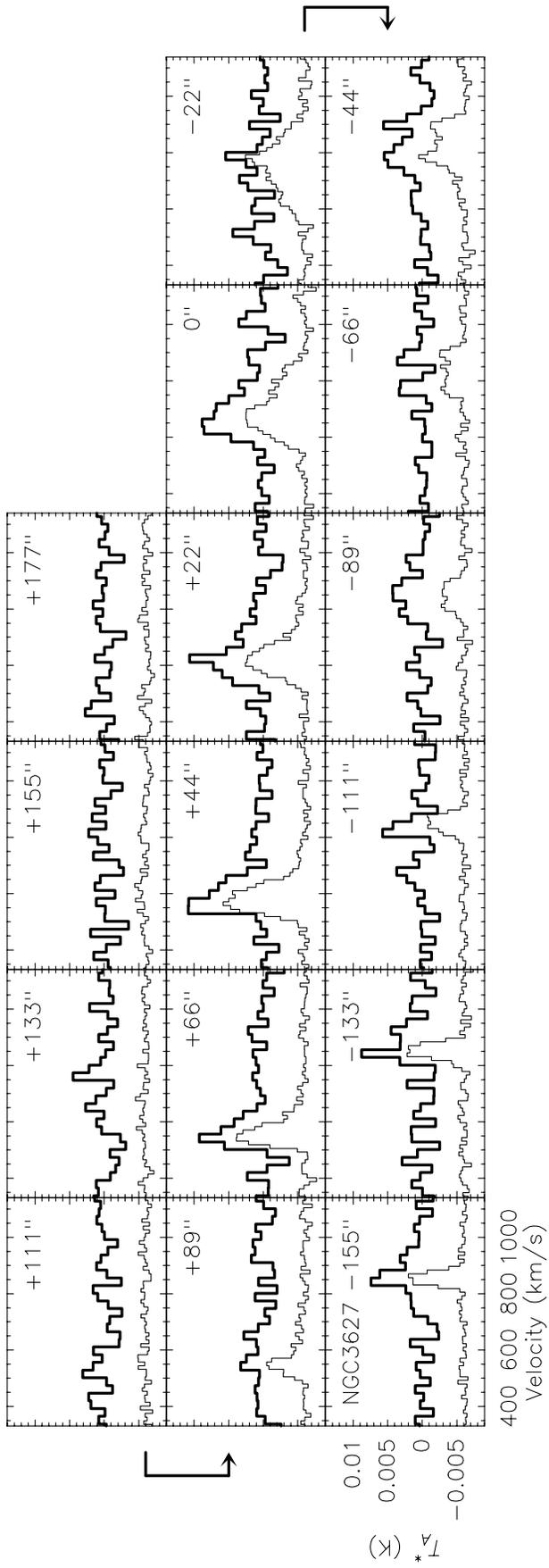
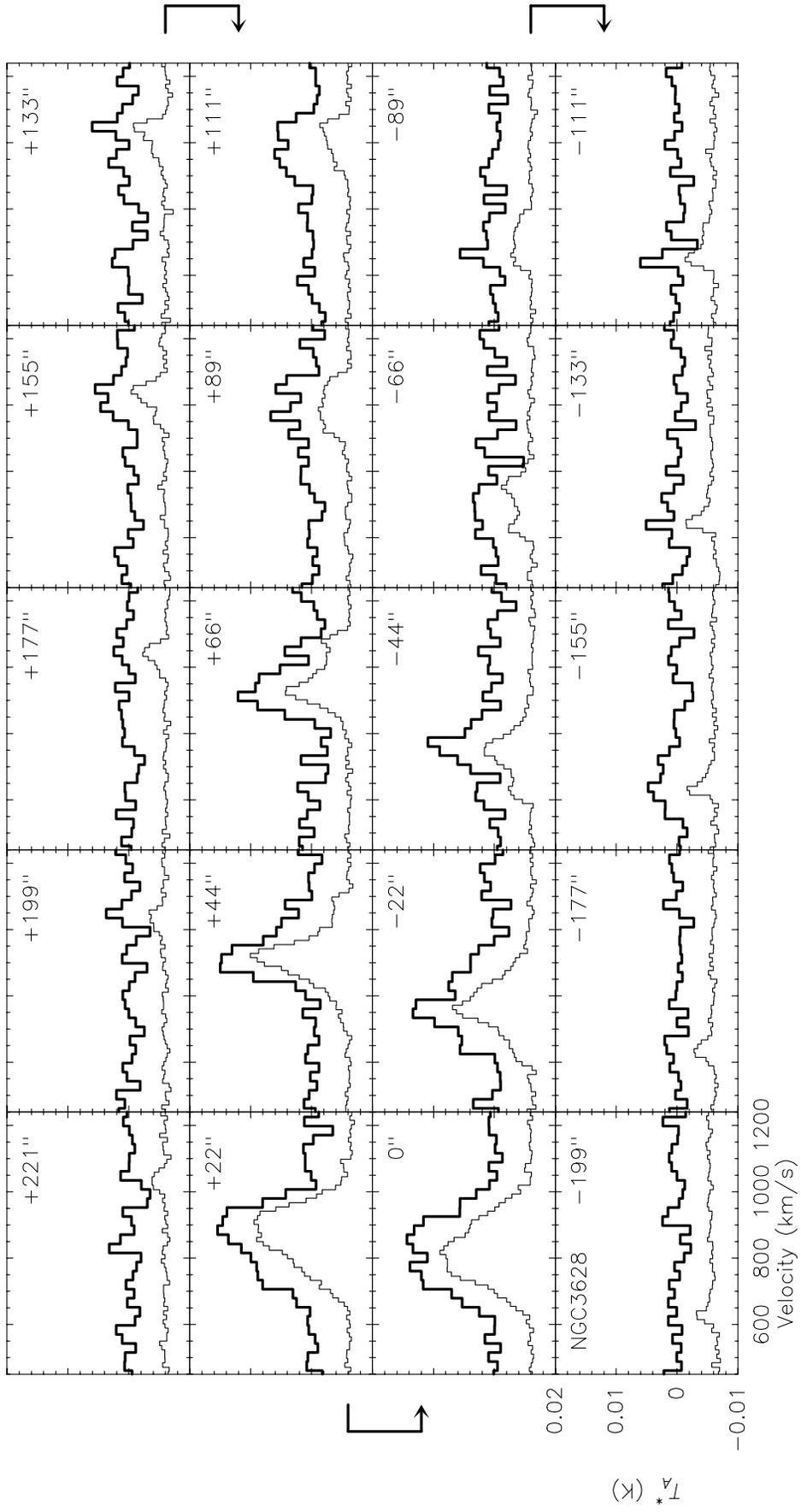
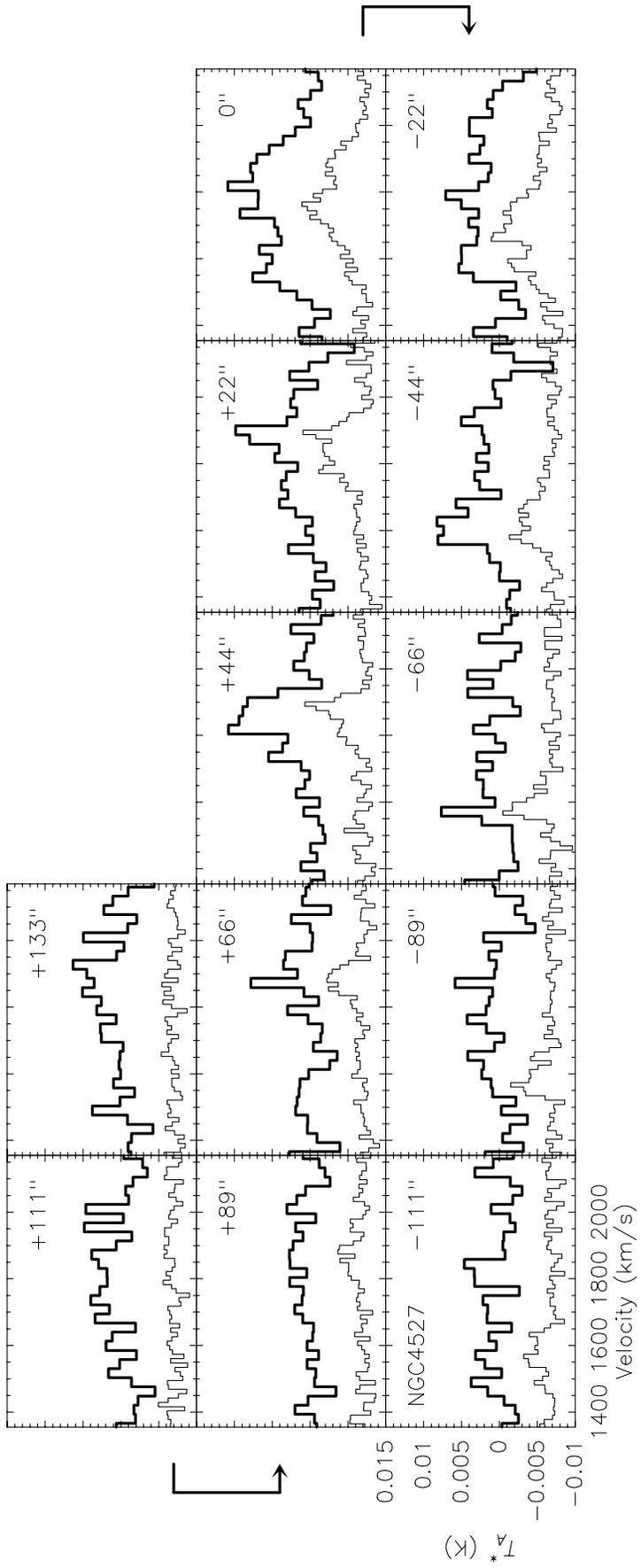
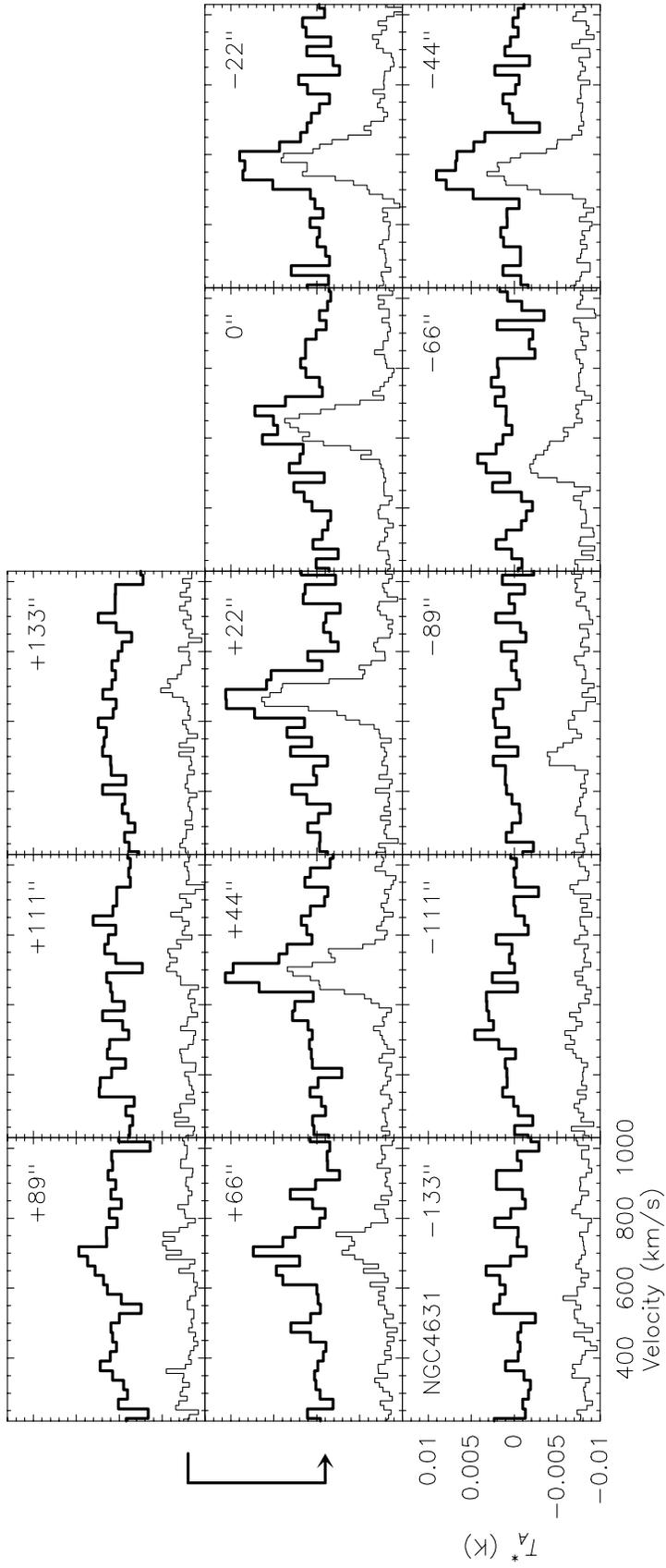
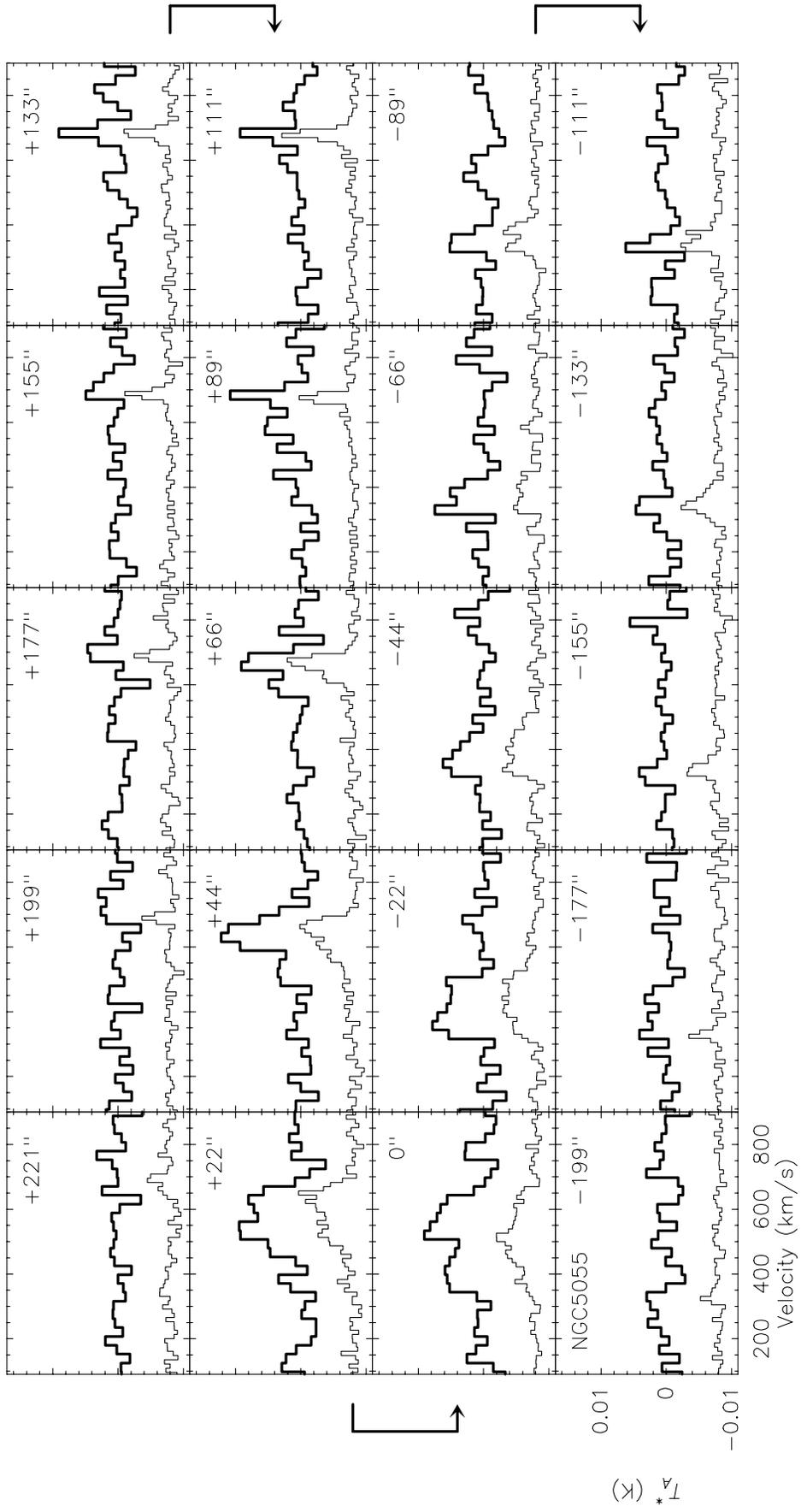
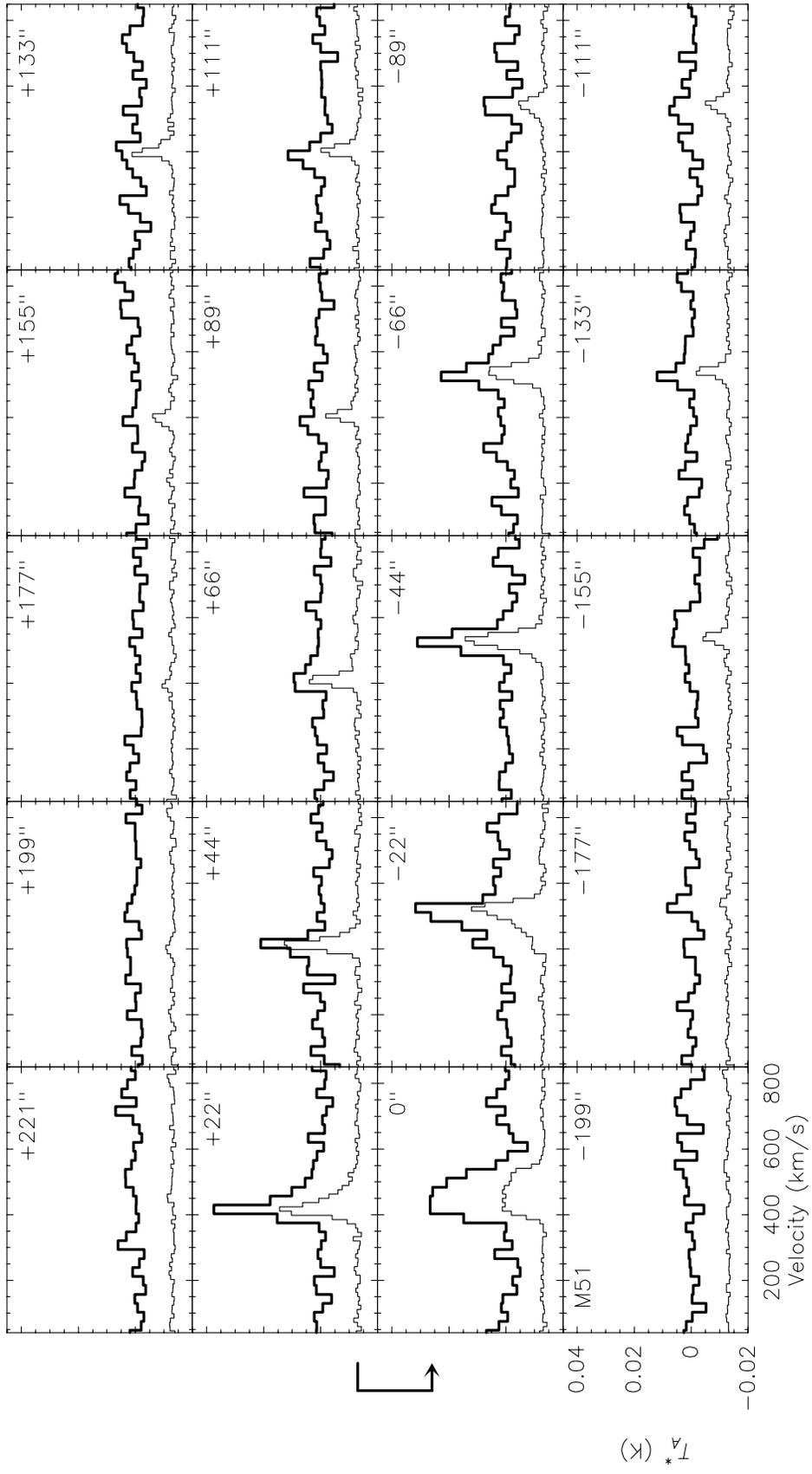
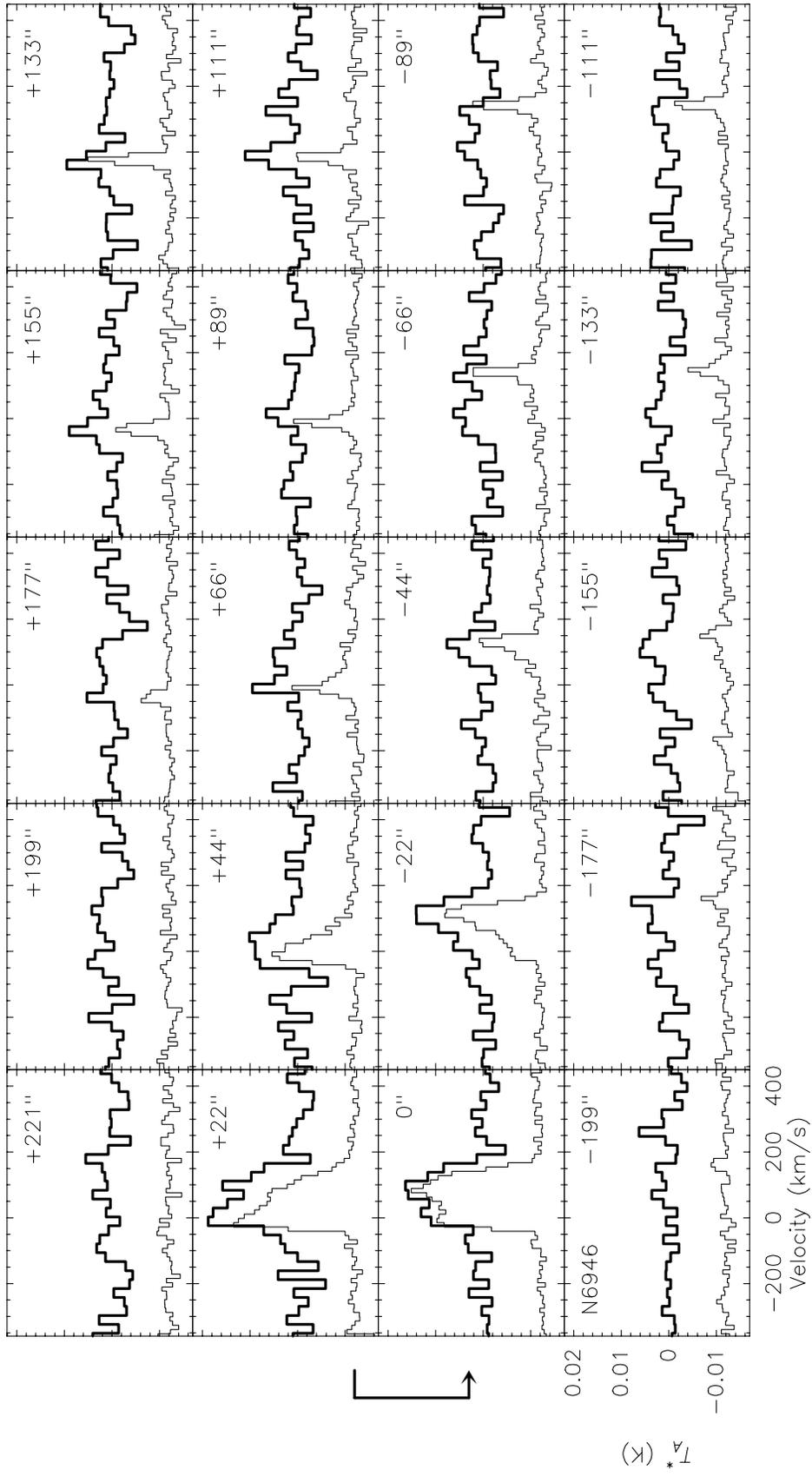
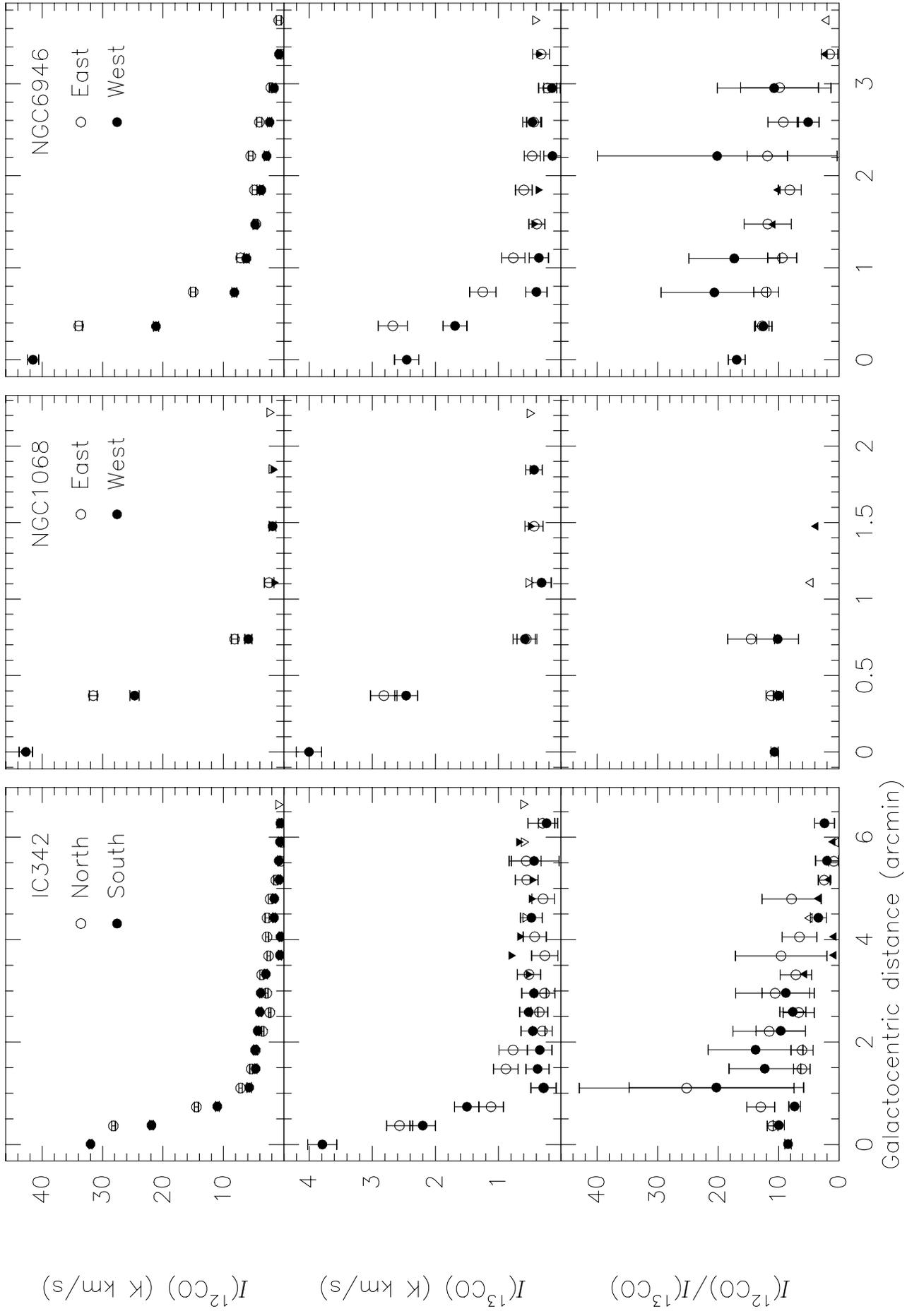
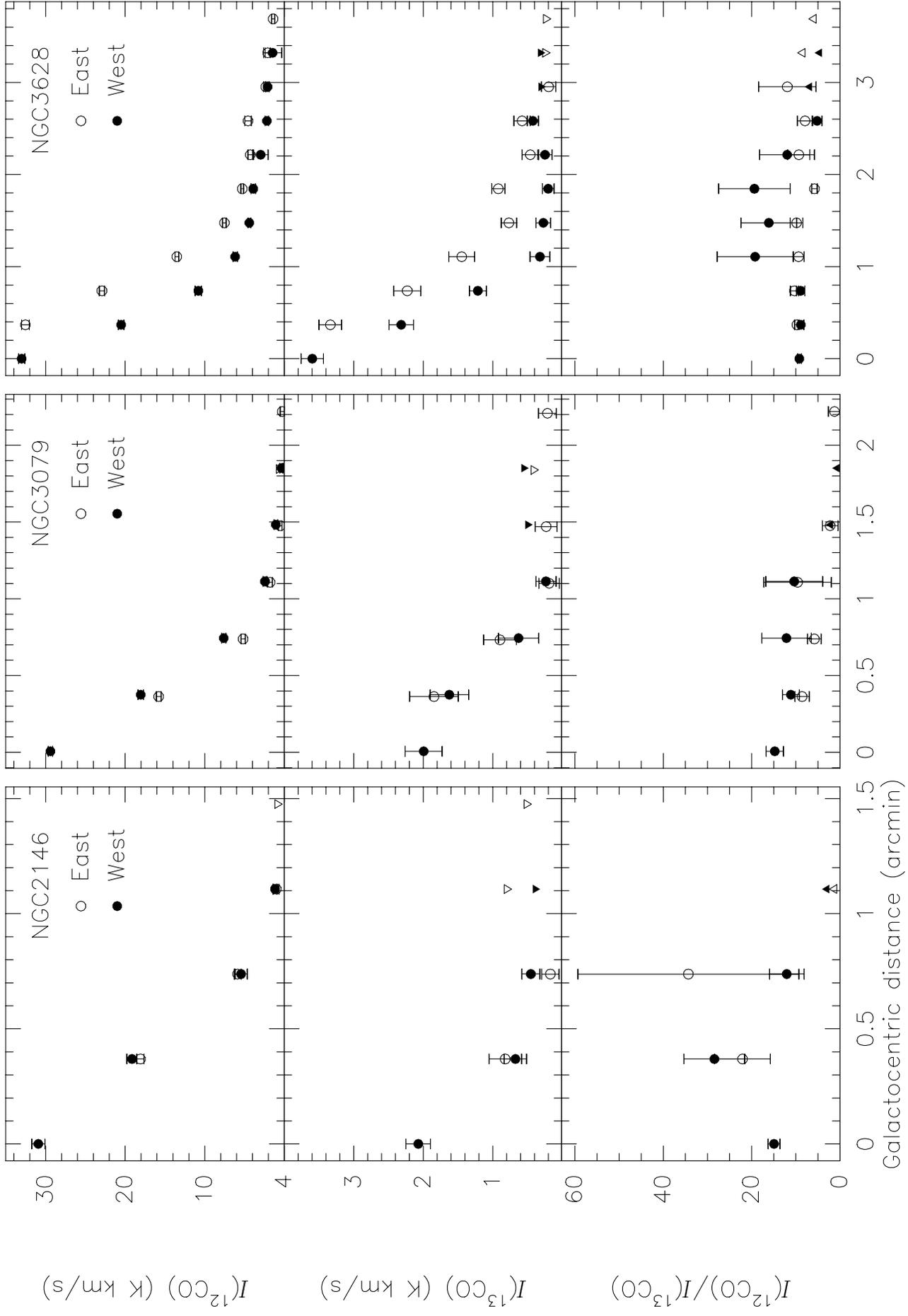
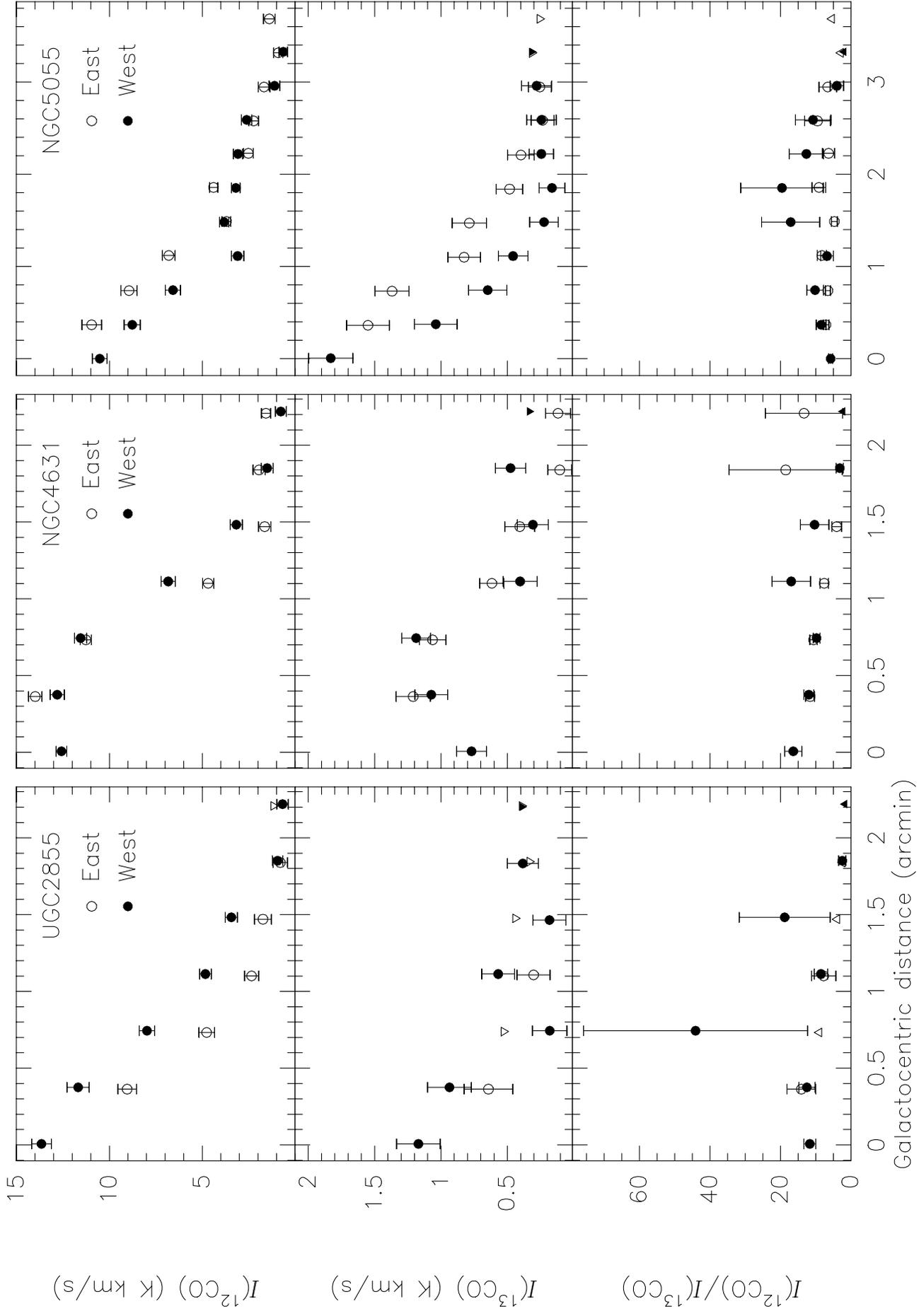
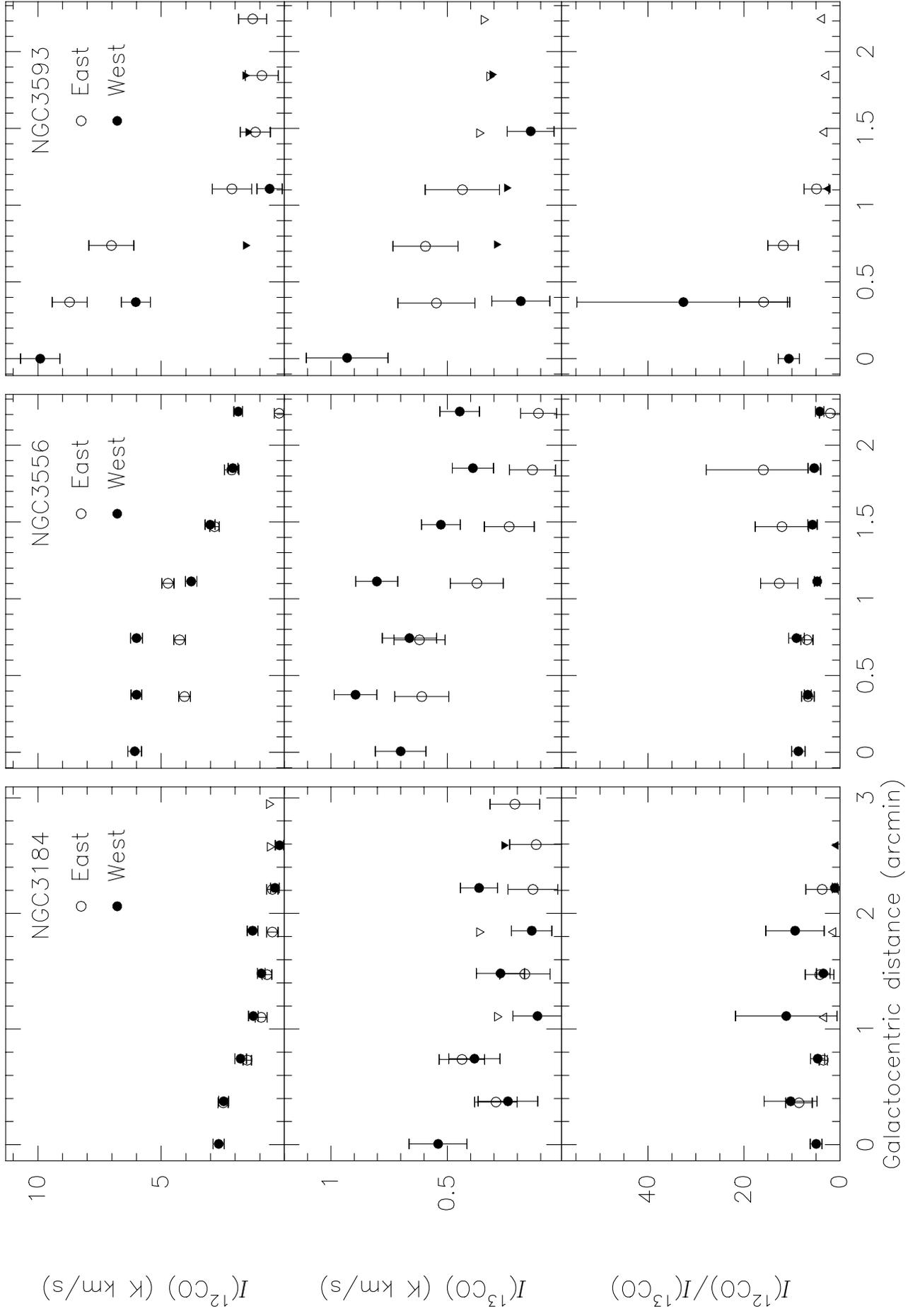
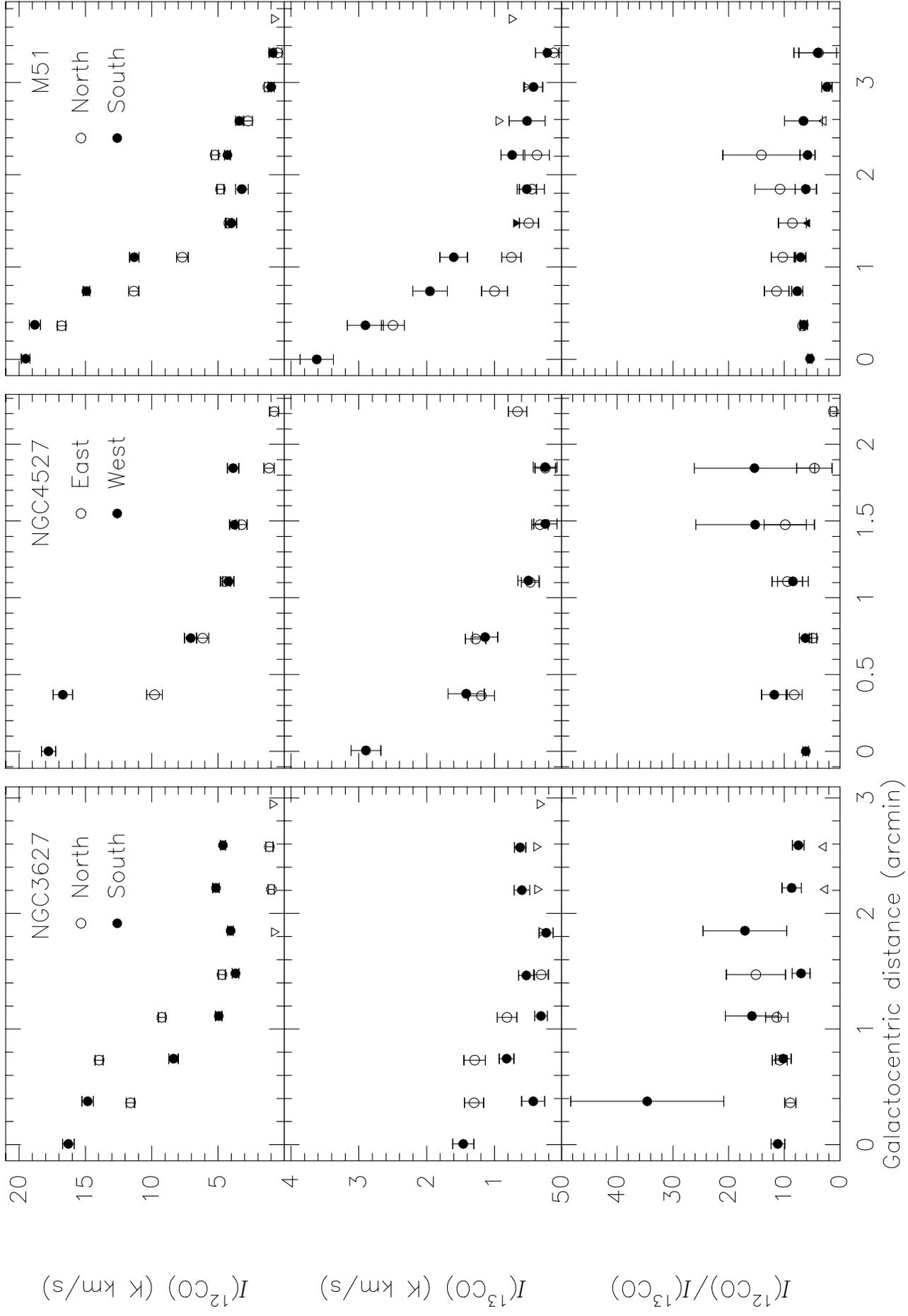
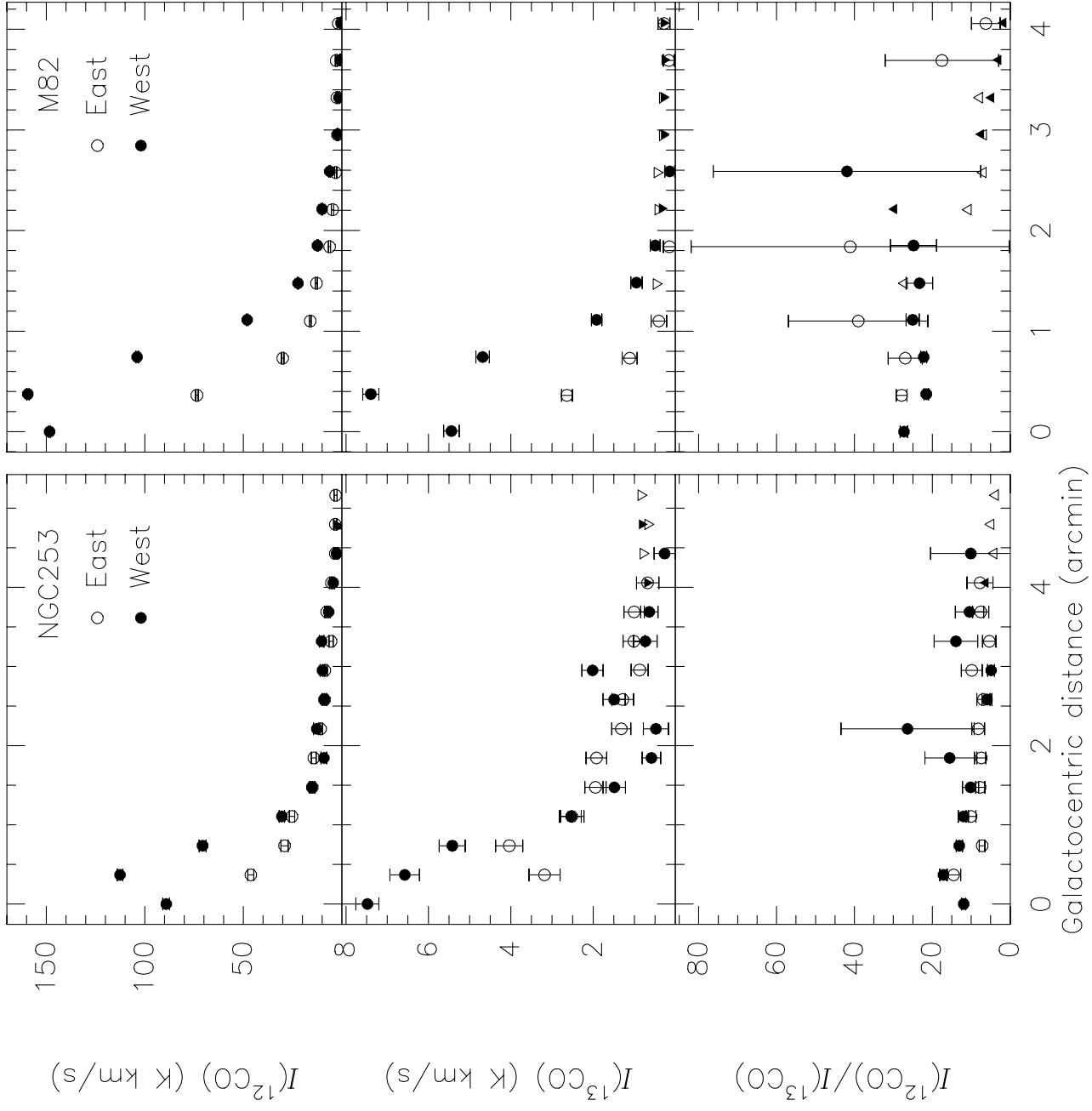
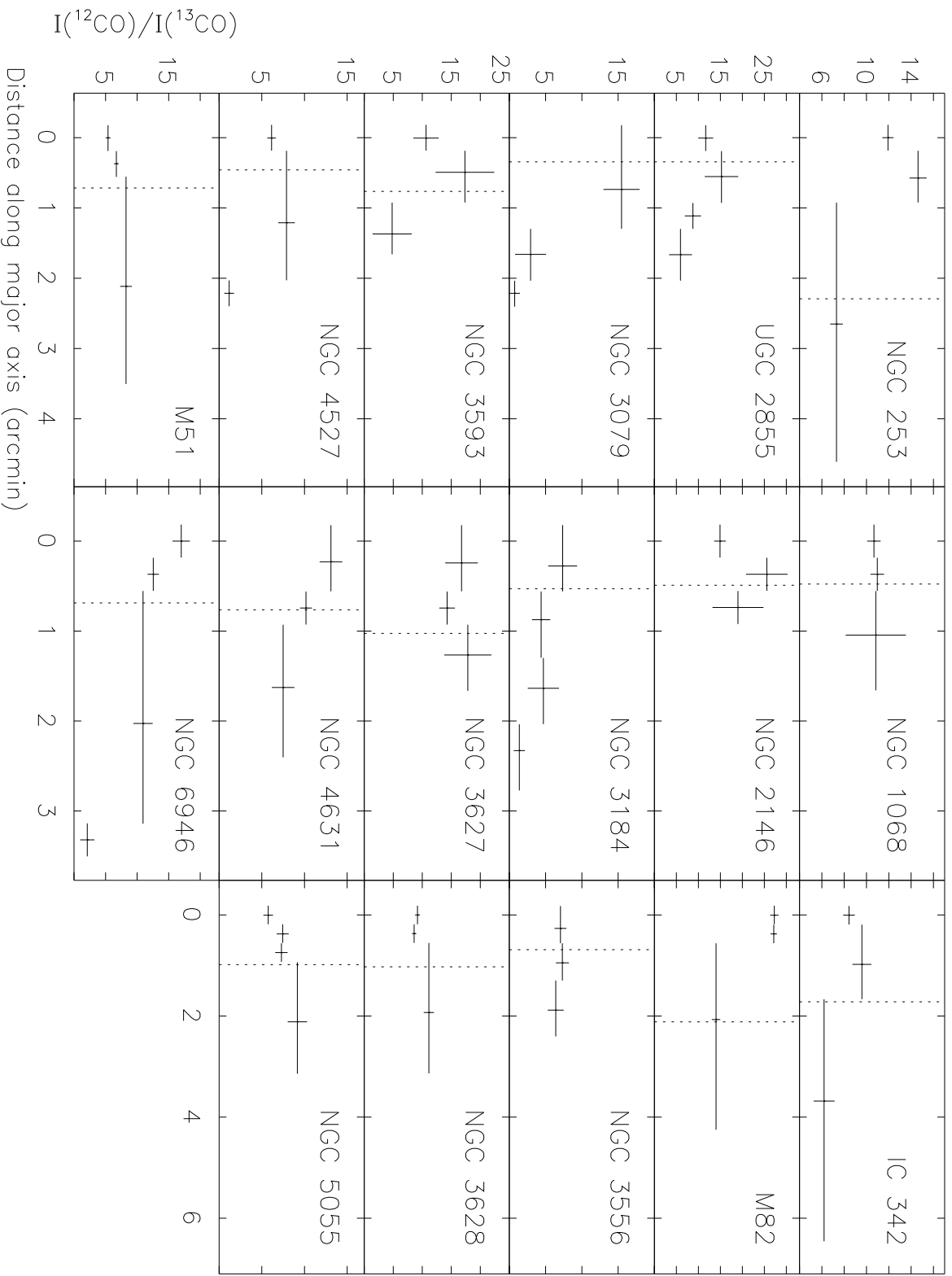
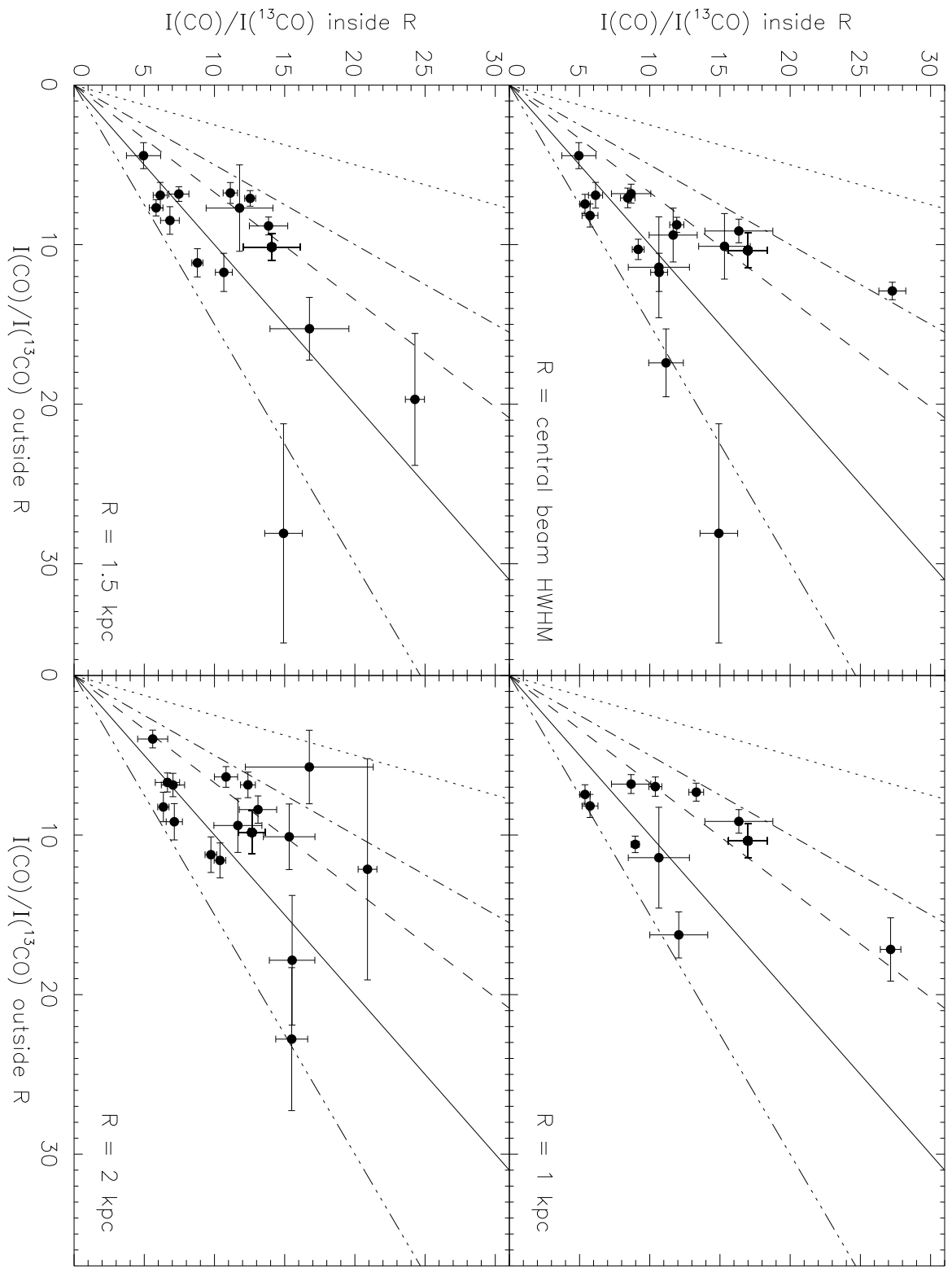
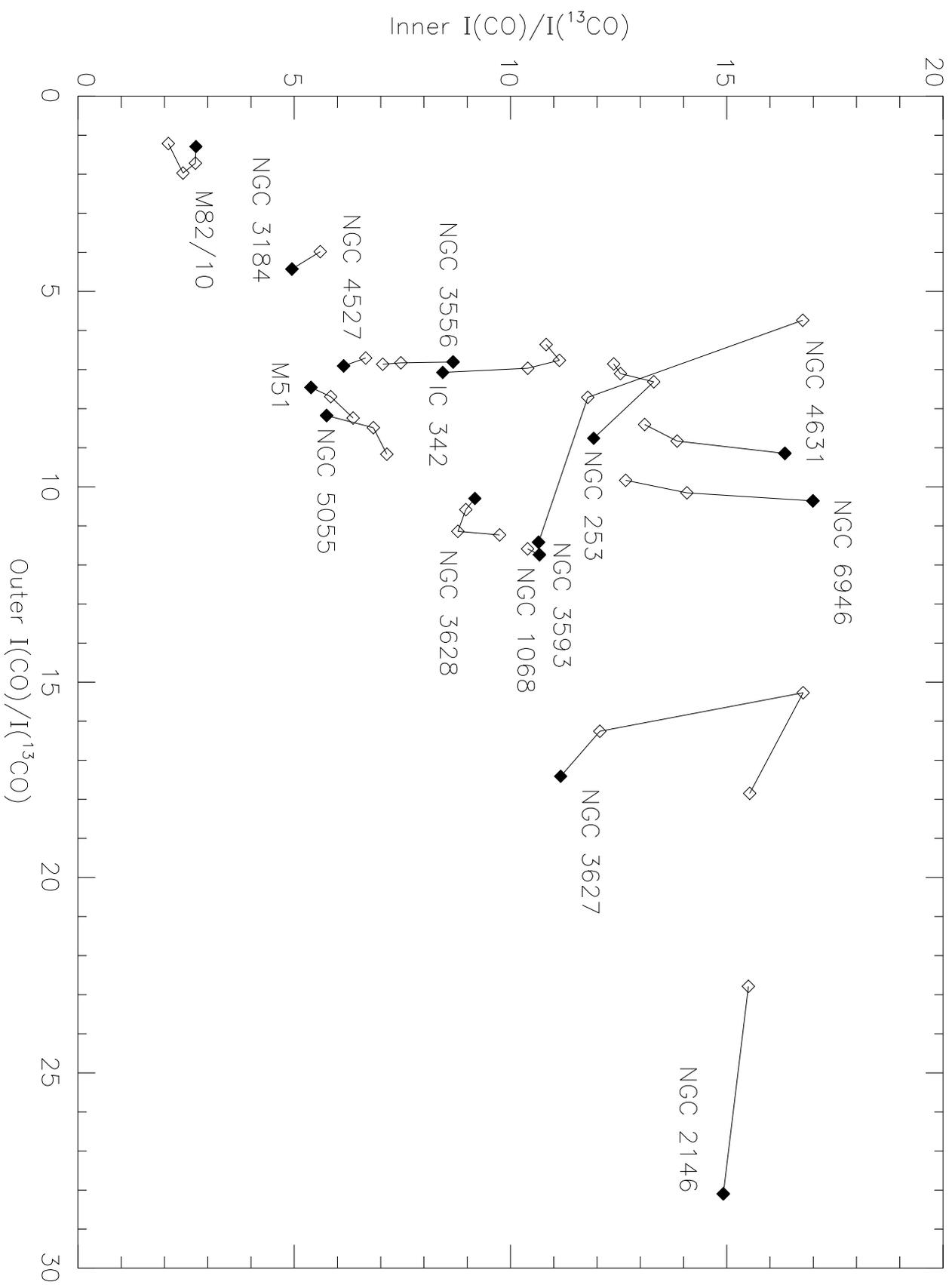
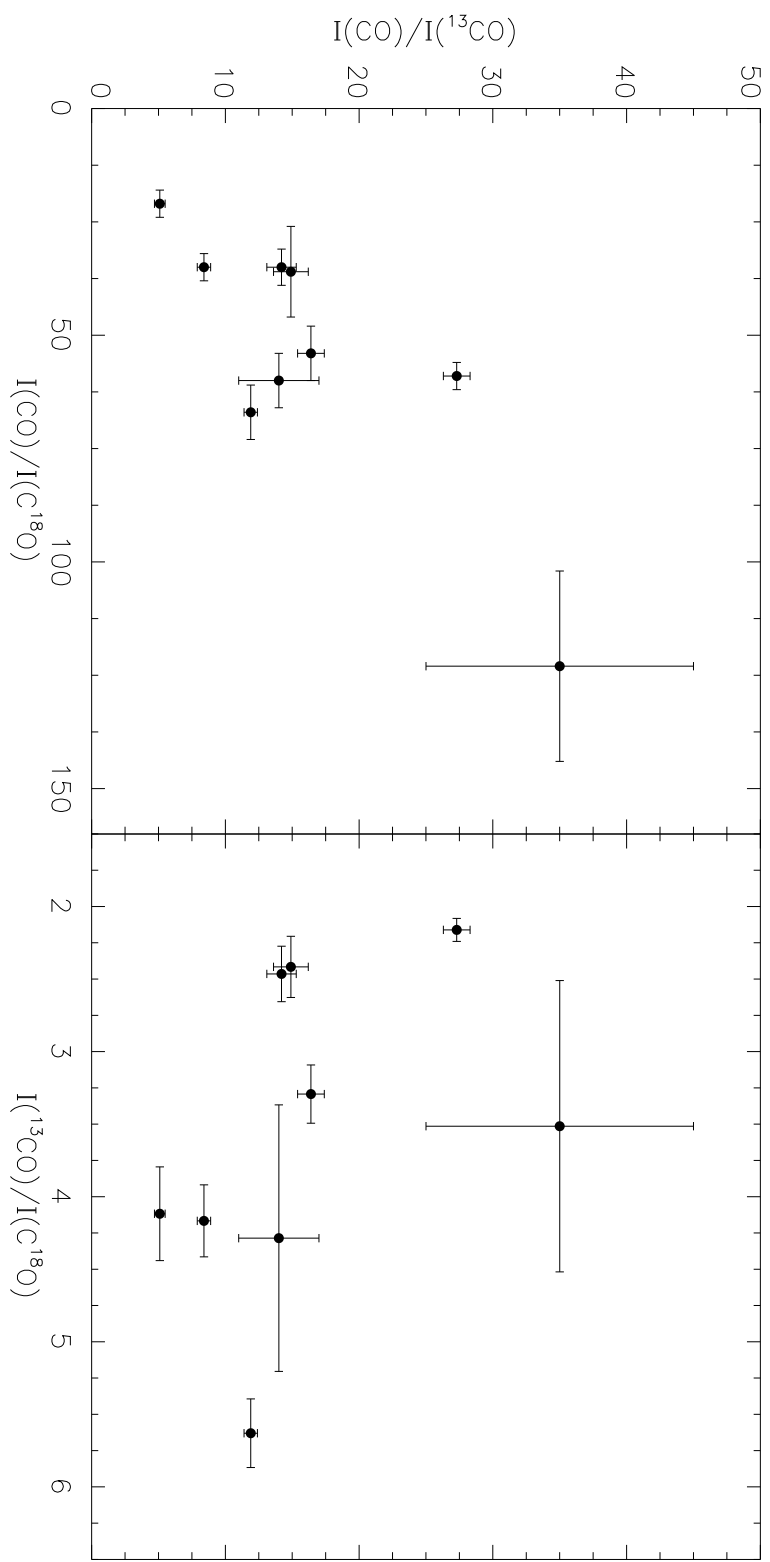
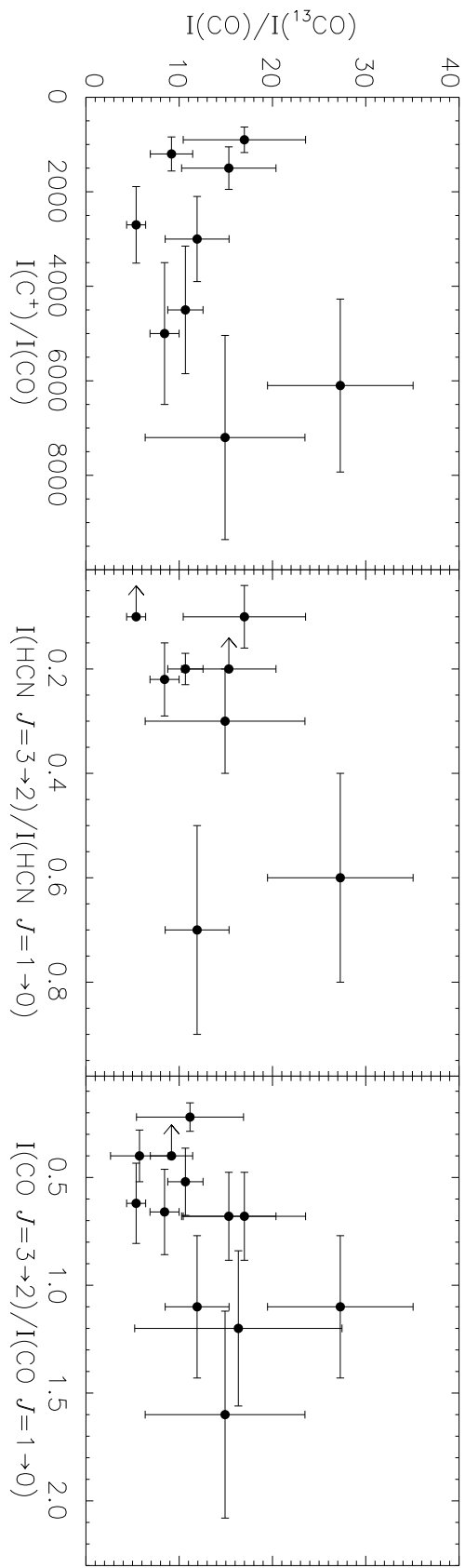
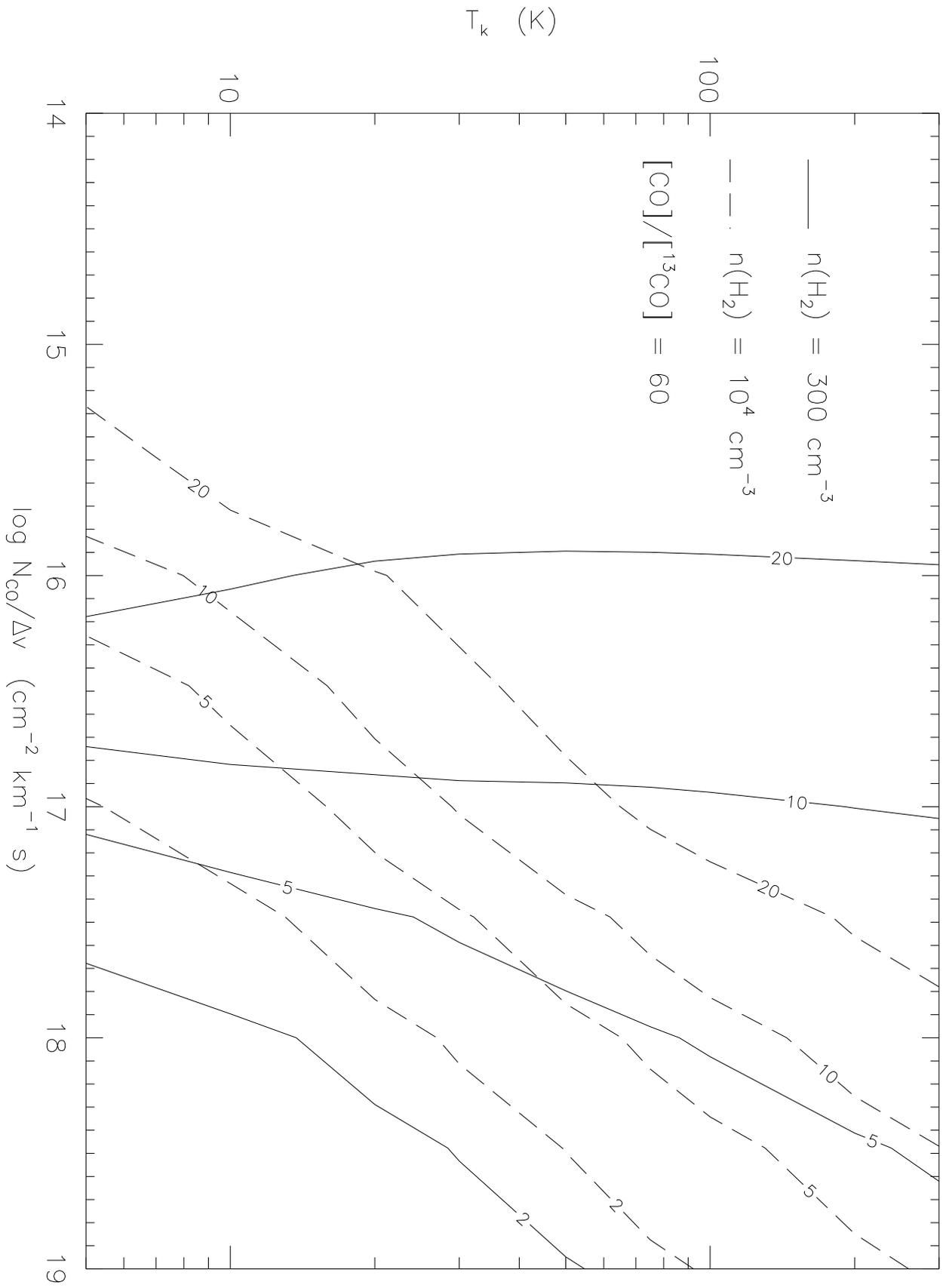
\end{document}